\documentclass[aps,prd,10pt,twocolumn,showkeys]{revtex4-2}

\PassOptionsToPackage{
  colorlinks=true,       % Links coloridos (azul) em vez de caixas
  linkcolor=blue,        % Cor para links internos
  citecolor=magenta,     % Cor para citações
  urlcolor=cyan,         % Cor para URLs
  pdftitle={Título do Artigo}, % Metadados do PDF
  pdfauthor={Autor}, 
  pdfencoding=auto,      % Suporte a caracteres especiais
  bookmarks=true,        % Gera bookmarks no PDF
  hypertexnames=false    % Evita conflitos com revtex
}{hyperref}

\usepackage{amsmath}           % Core mathematical typesetting (load first)
\usepackage{amssymb}           % Extended mathematical symbols
\usepackage{nccmath, mathtools}  % Medium-sized math commands (e.g., \medmath for matrix scaling)
\usepackage{bm}                % Bold mathematical symbols (\bm command)

\usepackage{ragged2e}          % Improved text alignment controls
\usepackage{soul}              % Text highlighting/decoration (e.g., \hl, \st)

\usepackage{graphicx}          % Image inclusion (already loaded by revtex, but safe to reload)
\graphicspath{{./}}            % Image search path
\usepackage[dvipsnames,table]{xcolor} % Color management (table option for colored tables)

\usepackage{dcolumn}           % Decimal alignment in tables (D column type)

\usepackage{listings}          % Typesetting source code

\usepackage{bbold}             % Blackboard bold symbols alternative (\mathbb{1} etc.)

\newcommand{\adj}{\mathrm{adj}} % Adjoint operator definition

\providecommand{\be}{ \begin{equation} }
\providecommand{\ee}{\end{equation}}
\providecommand{\bea}{\begin{eqnarray}}
\providecommand{\eea}{\end{eqnarray}}

\def\331{$SU(3)_C\otimes SU(3)_L\otimes U(1)_X$}
\def\3311{$\text{SU}(3)_C \otimes\text{SU}(3)_L\otimes\text{U}(1)_X \otimes \text{U}(1)_N$}

\definecolor{darkpastelpurple}{rgb}{0.59, 0.44, 0.84}
\definecolor{frenchlilac}{rgb}{0.53, 0.38, 0.56}
\definecolor{violet}{rgb}{0.56, 0.0, 1.0}
\definecolor{grey}{cmyk}{0,0,0,0.75}
\definecolor{tangerine}{cmyk}{0,0.5,1,0}
\definecolor{darkgreen}{cmyk}{1,0,1,0.23}
\definecolor{Red}{rgb}{1,0,0}
\definecolor{Blue}{rgb}{0,0,1}
\definecolor{Green}{rgb}{0,1,0}
\definecolor{Grey}{cmyk}{0,0,0,0.75}
\definecolor{Tangerine}{cmyk}{0,0.5,1,0}
\definecolor{Darkgreen}{cmyk}{1,0,1,0.23}
\definecolor{Cyan}{cmyk}{1,0,0,0}
\definecolor{Yellow}{cmyk}{0,0,1,0}
\definecolor{darkblue}{cmyk}{1,0.69,0,0.11}

\newcolumntype{s}{>{\columncolor[HTML]{ffebcd}}c}
\newcolumntype{q}{>{\columncolor[HTML]{fff4e3}}c}
\definecolor{exotic}{HTML}{c41525}
\definecolor{let}{HTML}{764b36}

\usepackage{tikz}
\usetikzlibrary{decorations.pathmorphing,decorations.markings}

\tikzset{
  psi/.style={
    decoration={
      markings,
      mark=at position 0.6 with {\arrow{>}}
    },
    postaction={decorate},
    double,
    double distance=1pt
  },
  psiNoArrow/.style={
    decoration={
      markings,
      mark=at position 0.6 with 
    },
    postaction={decorate},
    double,
    double distance=1pt
  },
  nucleon/.style={
    decoration={
      markings,
      mark=at position 0.6 with {\arrow{>}}
    },
    postaction={decorate}
  },
  external/.style={},  
  gluon/.style={
  decorate, draw=black, 
  decoration={coil,amplitude=4pt, segment length=5pt}
  },
  particle/.style={draw=black, postaction={decorate}, decoration={markings,mark=at position .5 with {\arrow[draw=black]{>}}}},
 photon/.style={decorate, decoration={snake,amplitude=2pt, segment length=5pt}, draw=black}
}
\newcommand{\dominantScalar}{
\begin{tikzpicture}[thick,scale=1.0]
    \draw[particle] (0,0) -- (1,0);
    \draw[particle] (1,0) -- (2.5,0);
    \draw[particle] (2.5,0) -- (4,0);
    \draw[particle] (4,0) -- (5,0);
    \draw[dashed]  (1,0) arc (180:0:1.5cm) ;
    \draw[photon,green] (2.5,0) -- (4,-1.3);

    \fill[black] (1,0) circle (0.06cm);
    \fill[black] (2.5,0) circle (0.06cm);
    \fill[black] (4,0) circle (0.06cm);

    \node at (0.5, -0.3) [external]{$\mu$};
    \node at (1.75, 0.3) [external]{$E'_i$};
    \node at (3.25, 0.3) [external]{$E'_i$};
    \node at (4.5, -0.3) [external]{$e$};
    \node at (2.5, 1.8) [external]{$H_3$};
    \node at (4.15,-1.25) [external] {$\gamma$};
\end{tikzpicture}
}

\newcommand{\dominantGauge}{
\begin{tikzpicture}[thick,scale=1.0]
    \draw[particle] (0,0) -- (1,0);
    \draw[particle] (1,0) -- (2.5,0);
    \draw[particle] (2.5,0) -- (4,0);
    \draw[particle] (4,0) -- (5,0);
    \draw[photon]  (1,0) arc (180:0:1.5cm) ;
    \draw[photon,green] (2.5,0) -- (4,-1.3);

    \fill[black] (1,0) circle (0.06cm);
    \fill[black] (2.5,0) circle (0.06cm);
    \fill[black] (4,0) circle (0.06cm);

    \node at (0.5, -0.3) [external]{$\mu$};
    \node at (1.75, 0.3) [external]{$e,\mu$};
    \node at (3.25, 0.3) [external]{$e,\mu$};
    \node at (4.5, -0.3) [external]{$e$};
    \node at (2.5, 1.8) [external]{$Z$};
    \node at (4.15,-1.25) [external] {$\gamma$};
\end{tikzpicture}
}

\newcommand{\ComboOne}{
\begin{tikzpicture}[thick,scale=1.0]
    \draw[particle] (0,0) -- (1,0);
    \draw[particle] (1,0) -- (2.5,0);
    \draw[particle] (2.5,0) -- (4,0);
    \draw[particle] (4,0) -- (5,0);
    \draw[dashed]  (1,0) arc (180:0:1.5cm) ;
    \draw[photon,green] (2.5,0) -- (4,-1.3);

    \fill[black] (1,0) circle (0.06cm);
    \fill[black] (2.5,0) circle (0.06cm);
    \fill[black] (4,0) circle (0.06cm);

    \node at (0.5, -0.3) [external]{$\mu$};
    \node at (1.75, 0.3) [external]{$E'_i$};
    \node at (3.25, 0.3) [external]{$E'_i$};
    \node at (4.5, -0.3) [external]{$e$};
    \node at (2.5, 1.8) [external]{$H_1,H_2,H_4$};
    \node at (4.15,-1.25) [external] {$\gamma$};
\end{tikzpicture}
}

\newcommand{\ComboTwo}{
\begin{tikzpicture}[thick,scale=1.0]
    \draw[particle] (0,0) -- (1,0);
    \draw[particle] (1,0) -- (2.5,0);
    \draw[particle] (2.5,0) -- (4,0);
    \draw[particle] (4,0) -- (5,0);
    \draw[dashed]  (1,0) arc (180:0:1.5cm) ;
    \draw[photon,green] (2.5,0) -- (4,-1.3);

    \fill[black] (1,0) circle (0.06cm);
    \fill[black] (2.5,0) circle (0.06cm);
    \fill[black] (4,0) circle (0.06cm);

    \node at (0.5, -0.3) [external]{$\mu$};
    \node at (1.75, 0.3) [external]{$E_i$};
    \node at (3.25, 0.3) [external]{$E_i$};
    \node at (4.5, -0.3) [external]{$e$};
    \node at (2.5, 1.8) [external]{$S_d$};
    \node at (4.15,-1.25) [external] {$\gamma$};
\end{tikzpicture}
}

\newcommand{\ComboThree}{
\begin{tikzpicture}[thick,scale=1.0]
    \draw[particle] (0,0) -- (1,0);
    \draw[particle] (1,0) -- (2.5,0);
    \draw[particle] (2.5,0) -- (4,0);
    \draw[particle] (4,0) -- (5,0);
    \draw[photon]  (1,0) arc (180:0:1.5cm) ;
    \draw[photon,green] (2.5,0) -- (4,-1.3);

    \fill[black] (1,0) circle (0.06cm);
    \fill[black] (2.5,0) circle (0.06cm);
    \fill[black] (4,0) circle (0.06cm);

    \node at (0.5, -0.3) [external]{$\mu$};
    \node at (1.75, 0.3) [external]{$\tau, E'_i$};
    \node at (3.25, 0.3) [external]{$\tau, E'_i$};
    \node at (4.5, -0.3) [external]{$e$};
    \node at (2.5, 1.8) [external]{$Z,Z',Z''$};
    \node at (4.15,-1.25) [external] {$\gamma$};
\end{tikzpicture}
}

\newcommand{\ComboFour}{
\begin{tikzpicture}[thick,scale=1.0]
    \draw[particle] (0,0) -- (1,0);
    \draw[particle] (1,0) -- (2.5,0);
    \draw[particle] (2.5,0) -- (4,0);
    \draw[particle] (4,0) -- (5,0);
    \draw[photon]  (1,0) arc (180:0:1.5cm) ;
    \draw[photon,green] (2.5,0) -- (4,-1.3);

    \fill[black] (1,0) circle (0.06cm);
    \fill[black] (2.5,0) circle (0.06cm);
    \fill[black] (4,0) circle (0.06cm);

    \node at (0.5, -0.3) [external]{$\mu$};
    \node at (1.75, 0.3) [external]{$E_i$};
    \node at (3.25, 0.3) [external]{$E_i$};
    \node at (4.5, -0.3) [external]{$e$};
    \node at (2.5, 1.8) [external]{$V^0$};
    \node at (4.15,-1.25) [external] {$\gamma$};
\end{tikzpicture}
}

\newcommand{\ComboFive}{
\begin{tikzpicture}[thick,scale=1.0]
    \draw[particle] (0,0) -- (1,0);
    \draw[particle] (1,0) -- (4,0);
    \draw[particle] (4,0) -- (5,0);
    \draw[photon]  (1,0) arc (180:0:1.5cm) ;
    \draw[photon,green] (2.5,1.5) -- (4,2.5);

    \fill[black] (1,0) circle (0.06cm);
    \fill[black] (2.5,1.5) circle (0.06cm);
    \fill[black] (4,0) circle (0.06cm);

    \node at (0.5, -0.3) [external]{$\mu$};
    \node at (2.5, 0.3) [external]{$\nu_{i},N_{i}$};
    \node at (4.5, -0.3) [external]{$e$};
    \node at (1, 1.2) [external]{$W^-$};
    \node at (4.1, 1.2) [external]{$W^-$};
    \node at (4.1,2.5) [external] {$\gamma$};
\end{tikzpicture}
}

\newcommand{\ComboSix}{
\begin{tikzpicture}[thick,scale=1.0]
    \draw[particle] (0,0) -- (1,0);
    \draw[particle] (1,0) -- (4,0);
    \draw[particle] (4,0) -- (5,0);
    \draw[photon]  (1,0) arc (180:0:1.5cm) ;
    \draw[photon,green] (2.5,1.5) -- (4,2.5);

    \fill[black] (1,0) circle (0.06cm);
    \fill[black] (2.5,1.5) circle (0.06cm);
    \fill[black] (4,0) circle (0.06cm);

    \node at (0.5, -0.3) [external]{$\mu$};
    \node at (2.5, 0.35) [external]{$f, f'_1,f'_2,f'_3$};
    \node at (4.5, -0.3) [external]{$e$};
    \node at (1, 1.2) [external]{$V^-$};
    \node at (4.1, 1.2) [external]{$V^-$};
    \node at (4.1,2.5) [external] {$\gamma$};
\end{tikzpicture}
}

\newcommand{\DMVee}{
\begin{tikzpicture}[thick,scale=1.0]
    \draw[particle] (2,0) -- (0,0);
    \draw[particle] (0,2) -- (2,2);
    \draw[photon] (2,0) -- (2,2);
    \draw[particle] (2,2) -- (4,2);
    \draw[particle] (4,0) -- (2,0);

    \fill[black] (2,0) circle (0.06cm);
    \fill[black] (2,2) circle (0.06cm);

    \node at (-0.3, -0.3) [external]{$f_d$};
    \node at (-0.3, 2.3) [external]{$f_d$};
    \node at (2.5, 1) [external]{$V^\pm$};
    \node at (4.3, 2.3) [external]{$e^-$};
    \node at (4.3, -0.3) [external]{$e^+$};
\end{tikzpicture}
}

\newcommand{\DMHee}{
\begin{tikzpicture}[thick,scale=1.0]
    \draw[particle] (1,1) -- (0,0);
    \draw[particle] (0,2) -- (1,1);
    \draw[dashed] (1,1) -- (3,1);
    \draw[particle] (3,1) -- (4,2);
    \draw[particle] (4,0) -- (3,1);

    \fill[black] (1,1) circle (0.06cm);
    \fill[black] (3,1) circle (0.06cm);

    \node at (-0.3, -0.3) [external]{$f_d$};
    \node at (-0.3, 2.3) [external]{$f_d$};
    \node at (2, 1.3) [external]{$H_1, H_2$};
    \node at (4.3, 2.3) [external]{$e^-, b$};
    \node at (4.3, -0.3) [external]{$e^+, \overline{b}$};
\end{tikzpicture}
}

\newcommand{\DMHWW}{
\begin{tikzpicture}[thick,scale=1.0]
    \draw[particle] (1,1) -- (0,0);
    \draw[particle] (0,2) -- (1,1);
    \draw[dashed] (1,1) -- (3,1);
    \draw[photon] (3,1) -- (4,2);
    \draw[photon] (4,0) -- (3,1);

    \fill[black] (1,1) circle (0.06cm);
    \fill[black] (3,1) circle (0.06cm);

    \node at (-0.3, -0.3) [external]{$f_d$};
    \node at (-0.3, 2.3) [external]{$f_d$};
    \node at (2, 1.3) [external]{$H_1, H_2$};
    \node at (4.3, 2.3) [external]{$W^-$};
    \node at (4.3, -0.3) [external]{$W^+$};
\end{tikzpicture}
}

\newcommand{\DMZWW}{
\begin{tikzpicture}[thick,scale=1.0]
    \draw[particle] (1,1) -- (0,0);
    \draw[particle] (0,2) -- (1,1);
        \draw[photon] (1,1) -- (3,1);
    \draw[photon] (3,1) -- (4,2);
    \draw[photon] (4,0) -- (3,1);

    \fill[black] (1,1) circle (0.06cm);
    \fill[black] (3,1) circle (0.06cm);

    \node at (-0.3, -0.3) [external]{$f_d$};
    \node at (-0.3, 2.3) [external]{$f_d$};
    \node at (2, 1.3) [external]{$Z$};
    \node at (4.3, 2.3) [external]{$W^-$};
    \node at (4.3, -0.3) [external]{$W^+$};
\end{tikzpicture}
}

\newcommand{\hc}{\text{h.c.}}

\newcommand{\abs}[1]{\left\lvert #1 \right\rvert}

\begin{document}

\lstset{frame=tb,
  	language=Matlab,
  	aboveskip=3mm,
  	belowskip=3mm,
 	showstringspaces=false,
	columns=flexible,
  	basicstyle={\small\ttfamily},
  	numbers=none,
  	numberstyle=\tiny\color{gray},
 	keywordstyle=\color{blue},
	commentstyle=\color{green},
  	stringstyle=\color{mauve},
  	breaklines=true,
  	breakatwhitespace=true
  	tabsize=3
}

\title{Dark matter in the scale-invariant 3-3-1-1 model}
\author{Alex G. Dias}
\affiliation{Centro de Ci\^encias Naturais e Humanas, Universidade Federal do ABC,\\
09210-580, Santo Andr\'e-SP, Brasil}
\author{Kristjan Kannike}
\email{Kristjan.Kannike@cern.ch}
\affiliation{National Institute of Chemical Physics and Biophysics, R\"{a}vala 10, 10143, Tallinn, Estonia}
\author{Niko Koivunen}
\email{niko.koivunen@kbfi.ee}
\affiliation{National Institute of Chemical Physics and Biophysics, R\"{a}vala 10, 10143, Tallinn, Estonia}
\author{Julio Leite}
\email{julio.leite@ific.uv.es}
\affiliation{AHEP Group, Institut de F\'{i}sica Corpuscular --
  C.S.I.C./Universitat de Val\`{e}ncia, Parc Cient\'ific de Paterna.\\
 C/ Catedr\'atico Jos\'e Beltr\'an, 2 E-46980 Paterna (Valencia) - Spain}
\author{Vinicius Padovani}
\email{vinicius.padovani.physics@gmail.com}
\affiliation{Centro de Ci\^encias Naturais e Humanas, Universidade Federal do ABC,\\
09210-580, Santo Andr\'e-SP, Brasil}
\author{B. L. S\'anchez-Vega}
\email{bruce@fisica.ufmg.br}
\affiliation{Departamento de F\'isica, UFMG, Belo Horizonte, MG 31270-901, Brasil.\label{addr1}}

\date{\today}

\begin{abstract}

We propose a novel scale-invariant model with the 3-3-1-1 gauge symmetry featuring a universal see-saw mechanism for all fermion masses, which, through the inclusion of additional vector-like quarks, provides a partial explanation for the observed fermion mass hierarchies. A discrete remnant of the gauge group, the matter parity ($P_M$), stabilises a fermionic dark matter candidate, and the scalar sector includes two triplets (minimal for 3-3-1 breaking) and two scalar singlets.

We identify the lightest $ P_M $-odd fermion, $ f_\mathrm{d} $, as a viable dark matter candidate. Our analysis shows that $f_\mathrm{d}$ satisfies the observed relic density constraint within the mass range $ 220 \, \textrm{GeV} \lesssim m_{f_\mathrm{d}} \lesssim 555 \, \textrm{GeV} $, primarily due to resonant annihilation via the new scalar $H_2$. While this mass range depends on the symmetry-breaking scale $ v_\chi $, which has a lower bound of $ v_\chi \gtrsim 3.6 \, \textrm{TeV} $ from LEP constraints on the $ \rho_0 $ parameter, we adopt a more conservative lower bound of $v_\chi > 10 \, \textrm{TeV}$. This choice is made to ensure that the $Z'$ boson mass remains above approximately $ 4 \, \textrm{TeV} $, and is motivated by recent LHC results and future projections for $Z'$ boson searches, which provide more stringent constraints than previous bounds or those from the $\rho_0$ parameter.

Spin-independent (SI) interactions dominate the direct detection phenomenology of $f_\mathrm{d}$. We calculate the SI elastic scattering cross-section and find that parameter points satisfying the relic density constraint are consistent with current experimental limits from LZ and PandaX-4T for certain parameter choices, particularly depending on the $\alpha_{12}$ angle. Some regions of the viable parameter space lie below the neutrino floor. Prospects for detection by future experiments like XLZD and PandaX-xT are also presented and discussed. 

\end{abstract}


\maketitle
\tableofcontents

\section{Introduction}

The enormous success of the Standard Model (SM) has still left unanswered quite a few questions, such as neutrino mass, dark matter (DM), and the number of fermion families. The last question finds an answer in the family of 3-3-1 models, which have been extensively studied in the literature~\cite{Singer:1980sw, Pisano:1991ee, Foot:1992rh, Frampton:1992wt, Montero:1992jk, Pleitez:1992xh, Foot:1994ym, Pleitez:1994pu, Ozer:1995xi, Hoang:1995vq}. One of the most notable features of these models is that the number of fermion families must be a multiple of three due to the non-trivial nature of anomaly cancellation. Additionally, these models exhibit a rich structure enabling the incorporation of new particles that can address other open problems of the SM, such as the dark matter puzzle, the strong CP problem, and the origin of neutrino masses~\cite{Montero:2011tg, Dong:2013wca, Dong:2015yra, Alves:2016fqe, Leite:2019grf, Leite:2020bnb, Montero:2017yvy}.

Motivated by the potential of 3-3-1 models to address these shortcomings, we propose a model with dynamical symmetry breaking based on the gauge group 
$\textrm{SU}(3)_{C} \otimes \textrm{SU}(3)_{L} \otimes \textrm{U}(1)_{X} \otimes \textrm{U}(1)_{N}$ that extends the 3-3-1 gauge group by an additional $U(1)$ symmetry \cite{Dong:2013wca,Dong:2015yra,Alves:2016fqe,Leite:2019grf,Leite:2020bnb}. The \(\textrm{SU}(3)_{C}\) represents the colour symmetry of the Standard Model, \(\textrm{SU}(3)_{L}\) extends the left-handed chirality symmetry, and \(\textrm{U}(1)_{X}\) and \(\textrm{U}(1)_{N}\) are a combination of charge and $B-L$.  Building on this symmetry framework, this work presents a new model that, while sharing the same gauge symmetry as the framework of~\cite{Dias:2022hbu}, departs significantly from it in both scalar and fermionic field content, as well as in phenomenological structure. The extended gauge sector leaves remnant symmetries after the spontaneous symmetry breaking, which allow us to economically implement the universal see-saw mechanism and to stabilize dark matter.

One of the driving forces behind the proposed model is the reduction in the number of scalar fields in high representations of the gauge group and the free parameters in the scalar potential, especially those with mass dimension, when compared to the original 3-3-1 models. These parameters tend to make the models less predictive and introduce scalar complexity that, in some cases, becomes excessive compared to other extensions of the SM, such as the two-Higgs-doublet model (see \cite{Branco:2011iw}) and the minimal supersymmetric standard model (see \cite{Martin:1997ns}). For instance, in the original 3-3-1 models, the scalar sector includes an enlarged set of non-trivial \(\textrm{SU}(3)_{L}\) multiplets of scalar fields, such as three triplets~\cite{Singer:1980sw, Frampton:1992wt, Montero:1992jk, Pleitez:1992xh, Foot:1994ym, Pleitez:1994pu, Ozer:1995xi, Hoang:1995vq} and, in some cases, three triplets plus a sextet~\cite{Foot:1992rh}.

In order to achieve a very compact and more predictive scalar sector, we follow the same strategy as in~\cite{Dias:2022hbu}. Specifically, we introduce the minimal number of triplets required to dynamically break the gauge symmetry to
$\textrm{SU}(3)_{C} \otimes \textrm{U}(1)_{Q}$ via the Coleman-Weinberg mechanism \cite{Coleman:1973jx} in the Gildener-Weinberg approximation \cite{Gildener:1976ih}. The minimal number of triplets needed to achieve this is two. However, as first discussed in~\cite{Montero:2011tg}, despite these two triplets breaking the gauge symmetry, accidental chiral symmetries typically remain. These symmetries are usually associated with generators that are combinations of the diagonal generators of \(\textrm{SU}(3)_{L}\) and a Peccei-Quinn-like symmetry, as shown in~\cite{Montero:2014uya}.

As it happens, this symmetry is also chiral in the quark sector (and occasionally in the lepton sector), leaving some particles massless. This situation is inconsistent with experimental observations. Therefore, the accidental symmetry must be broken to generate mass terms for these  particles. To address this issue, we introduce new vector-like quarks and leptons.

At first glance, this might seem like a mere translation of the complexity from the scalar sector to the fermion sector. However, there are several advantages to this approach. The first is that these new fermions can be used to implement a universal see-saw mechanism \cite{Davidson:1987mh}, which partially addresses the hierarchy of fermion masses—an unresolved issue in the Standard Model. In our construction, the see-saw truly deserves the qualifier “universal”: a single set of vector-like partners simultaneously generates predictive mass matrices for quarks, charged leptons, and neutrinos—whereas earlier 3-3-1 approaches handled each sector in isolation or relied on ad-hoc extensions.

Moreover, these new fermions acquire masses at an energy scale related to scale invariance, which in this model is significantly higher than the electroweak and 3-3-1 scales due to the requirements of the see-saw mechanism. As a result, the number of physical degrees of freedom at the TeV scale becomes smaller and more manageable, making the model more feasible for phenomenological studies.

As a benefit of the extended gauge group, the matter parity that stabilises DM arises as a discrete remnant symmetry from the symmetry breaking \cite{Dimopoulos:1981zb,Farrar:1978xj,Ibanez:1991pr,Kadastik:2009cu,Kadastik:2009dj,Dong:2013wca,Dong:2014wsa,Alves:2016fqe,Dias:2022hbu}. While we have several potential dark matter candidates, only a dark Majorana fermion $f_\mathrm{d}$ can satisfy the direct detection constraints. Its stability is protected by a residual $\mathbb{Z}_2$ symmetry that naturally remains unbroken due to the representation content and gauge charge assignments, without the need to impose any additional discrete symmetry. Furthermore, radiative corrections involving $f_\mathrm{d}$ contribute significantly to the stabilization of the scalar potential, establishing a non-trivial connection between the dark matter sector and vacuum dynamics.

We systematically explore the key aspects of the proposed model. In Section~\ref{sec:model}, we define the field representations, charge assignments, and symmetry-breaking pattern, emphasizing the role of the scalar triplets $\eta$ and $\chi$ in inducing a distinct vacuum structure. Section~\ref{scsect} is devoted to the scalar sector, where we analyze the full scalar potential and express the quartic couplings in terms of parameters such as scalar masses, VEVs, and mixing angles, which provide a more physically transparent and intuitive parametrization. In Section~\ref{sec:pert}, we derive perturbative unitarity constraints at tree level, which prove to be highly constraining: several regions of parameter space, especially those with large scalar mixings, are excluded due to violations of unitarity bounds.

Section~\ref{sec:gauge} provides a complete analysis of the gauge boson spectrum. We derive analytical expressions for the masses of the charged and neutral gauge bosons and perform the diagonalization of the neutral gauge boson mass matrix, which has a particularly nontrivial structure. This allows for the identification of the physical states and the isolation of the heavy neutral boson $Z''$, whose presence is essential to ensuring vacuum stability. 
Section~\ref{scYukawa} is devoted to the generation of fermion masses via a universal see-saw mechanism, elegantly yielding the observed hierarchical pattern across all fermionic sectors.

In Section~\ref{sec:bounds}, we examine phenomenological constraints from lepton flavour-violating processes. Section~\ref{sec:dm} is devoted to the dark matter sector, demonstrating that the model successfully reproduces the observed relic density and features parameter regions within the projected sensitivity of future experiments.
Our main results and conclusions are summarized in Section~\ref{sec:conc}.

\section{The Scale-invariant 3-3-1-1 Model} \label{sec:model}

Although we follow the same general strategy as in~\cite{Dias:2022hbu}, there are some important differences in both the symmetry structure and the field content. To be more specific, we choose to introduce the \(\eta\) triplet (instead of \(\rho = \left(\rho^{+}_{1},\, \rho^{0}_{2},\, \rho^{+}_{3}\right)^T \sim \left(1,\,3,\,2/3\right)\)) and the \(\chi\) triplet, which transform under the \(\textrm{SU}(3)_{C} \otimes \textrm{SU}(3)_{L} \otimes \textrm{U}(1)_{X} \otimes \textrm{U}(1)_{N}\) group as \((1,3,-1/3,1/3)\) and \((1,3,-1/3,-2/3)\), respectively.

In other words,
\begin{align}
\begin{aligned}
    \eta &= \left(\eta_1^0 \equiv \frac{v_\eta + S_{\eta_1} + i A_{\eta_1}}{\sqrt{2}},\, \eta^{-}_2,\, \eta_3^0 \right)^T, \quad \\
\chi &= \left( \chi_1^0,\, \chi_2^{-},\, \chi_3^0 \equiv \frac{v_\chi + S_{\chi_3} + i A_{\chi_3}}{\sqrt{2}} \right)^T.\label{eq:scalartriplets}  
\end{aligned}
\end{align}

Note that \(\eta\) and \(\chi\) transform differently under the \(\textrm{U}(1)_{N}\) symmetry. This is an important point because we require that \(\eta_1^0\) and \(\chi_3^0\) acquire vacuum expectation values (VEVs) in order to break the gauge symmetry down to \(\textrm{SU}(3)_{C} \otimes \textrm{U}(1)_{Q}\). If these fields transformed identically under all the gauge groups, a gauge rotation could be performed to set one of the VEVs to zero.

As usual, we also embed the SM quarks and leptons in the representations:
\begin{align}
\begin{aligned}
L_{iL} &= \left(\nu_{i},\,e_{i},\, f_{i}\right)_{L}^T, \quad 
Q_{aL} = \left(d_{a},\,-u_{a},\, D_{a}\right)_{L}^T, \\
Q_{3L} &= \left(u_{3},\,d_{3},\, U_{3} \right)_{L}^T, \label{eq:triplets}  
\end{aligned}
\end{align}
where \(a = 1,\,2\), \(i = 1,\,2,\,3\), and \(L_{iL}\), \(Q_{aL}\), and \(Q_{3L}\) transform under the gauge group \(\textrm{SU}(3)_{C} \otimes \textrm{SU}(3)_{L} \otimes \textrm{U}(1)_{X} \otimes \textrm{U}(1)_{N}\) as \( (1,3,-1/3,-2/3) \), \( (3,\bar{3},0,0) \), and \( (3,3,1/3,2/3) \), respectively. 
As previously mentioned, the two scalar triplets \(\eta\) and \(\chi\) in eq.~\eqref{eq:scalartriplets} are not enough to generate masses for all the quarks and leptons. Thus, we introduce vector-like fermions \(\mathcal{F}_{iL,R}\), \({\cal{K}}_{a\,L,R}\), and \({\cal{K}}_{3L,R}\):
\begin{equation}
\begin{split}
\mathcal{F}_{iL,R} &= \left({E_{i},\,-f^\prime_{i},\,E^\prime_{i}}\right)_{L,R}^T, \\
\mathcal{K}_{a\,L,R} &= \left({{\mathcal{U}}_{a},{\mathcal{D}}_{a},\,{\mathcal{U}}^\prime_{a}}\right)_{L,R}^T, \\
\mathcal{K}_{3L,R} &= \left({\mathcal{D}_{3},\,-\mathcal{U}_{3},\,\mathcal{D}^\prime_{3}}\right)_{L,R}^T,
\label{eq:vectorlikefermions}    
\end{split}
\end{equation}
in order to allow terms such as \(\overline{L_{iL}}\, \eta^* \,\mathcal{F}_{jR}\), \(\overline{\mathcal{F}_{iL}}\, \chi^* \,e_{jR}\), $\overline{Q_{aL}}\, \eta\, \mathcal{K}_{bR}$ and $\overline{Q_{3L}} \,\eta^*\, \mathcal{K}_{3R} $ which ensure that all the SM fermions become massive, the fields  $\mathcal{F}_{iL,R}$, $\mathcal{K}_{aL,R}$ and $\mathcal{K}_{3L,R}$ have to belong to \( (1,\bar{3},-2/3,-1/3)\), \((3,3,1/3,-1/3)\) and \((3,\bar{3},0,1)\) representation of the 3-3-1-1 gauge group. 

However, it is still necessary to introduce a new real singlet scalar, 
\begin{equation}
\varphi = \frac{v_\varphi + S_\varphi}{\sqrt{2}}\sim(1,1,0,0),
\label{eq:varphi}
\end{equation}
to give masses for the \(\mathcal{F}_{iL,R}\), \({\cal{K}}_{a\,L,R}\), and \({\cal{K}}_{3L,R}\) through terms such as  $\varphi\overline{\mathcal{F}_{iL}}\, \mathcal{F}_{jR}$, $\varphi\, \overline{\mathcal{K}_{aL}} \,\mathcal{K}_{bR}
$ and $ \varphi\, \overline{\mathcal{K}_{3L}} \,\mathcal{K}_{3R}$. 

Although the lepton triplet \(L_{iL}\) in eq.~\eqref{eq:triplets} contains \(\nu_{iL}\), the gauge symmetries do not allow neutrino masses as in the Standard Model. To address this, we introduce right-handed neutrinos \(\nu_{iR} \sim (1,1,0,-1)\), motivated by current oscillation experiments, which require at least two neutrinos to be massive and necessarily non-degenerate. Together with the introduction of a complex scalar field
\begin{equation}
    \sigma = \frac{v_\sigma + S_\sigma + i A_\sigma}{\sqrt{2}} \sim (1,1,0,2),
\end{equation}
this allows the implementation of a type-I see-saw mechanism \cite{Gell-Mann:1979kx,Yanagida:1979uq,Mohapatra:1979ia,Glashow:1979nm,Minkowski:1977sc}. Note that \(\sigma\) is the only field responsible for breaking the \(\textrm{U}(1)_{N}\) symmetry and for giving mass to the neutral gauge boson associated with that symmetry.

\begin{table*}[t]
\setlength{\tabcolsep}{7pt} 
\centering
\begin{tabular}{|c||c|c|c|c|}
\hline 
\textbf{Fields} & \textbf{3-3-1-1} & \textbf{Q} & \textbf{B$-$L} & \textbf{$P_M$-Parity} \\
\hline
$L_{iL}$ & $(1,3,-1/3,-2/3)$ & $(0,-1,0)^T$ & $(-1,-1,0)^T$ & $(+,+,-)^T$ \\
$e_{iR}$ & $(1,1,-1,-1)$ & $-1$  & $-1$ & $+$ \\
$Q_{aL}$ & $(3,\bar{3},0,0)$ & $(-1/3,2/3,-1/3)^T$ & $(1/3,1/3,-2/3)^T$ & $(+,+,-)^T$ \\
$Q_{3L}$ & $(3,3,1/3,2/3)$ & $(2/3,-1/3,2/3)^T$ & $(1/3,1/3,4/3)^T$ & $(+,+,-)^T$ \\ 
$u_{iR}$ & $(3,1,2/3,1/3)$ & $2/3$ & $1/3$ & $+$ \\
$d_{iR}$ & $(3,1,-1/3,1/3)$ & $-1/3$ & $1/3$ & $+$ \\
$\nu_{iR}$ & $(1,1,0,-1)$ & $0$  & $-1$ & $+$ \\
$f_{iR}$  & $(1,1,0,0)$ & $0$  & $0$ & $-$ \\
$\mathcal{F}_{iL,R}$ & $(1,\bar{3},-2/3,-1/3)$ & $(-1,0,-1)^T$ & $(0,0,-1)^T$ & $(-,-,+)^T$ \\
$U_{3R}$ & $(3,1,2/3,4/3)$ & $2/3$  & $4/3$ & $-$ \\
$D_{aR}$ & $(3,1,-1/3,-2/3)$  & $-1/3$  & $-2/3$ & $-$ \\
$\mathcal{K}_{aL,R}$ & $(3,3,1/3,-1/3)$ & $(2/3,-1/3,2/3)^T$ & $(-2/3,-2/3,1/3)^T$ & $(-,-,+)^T$ \\
$\mathcal{K}_{3L,R}$ & $(3,\bar{3},0,1)$ & $(-1/3,2/3,-1/3)^T$ & $(4/3,4/3,1/3)^T$ & $(-,-,+)^T$ \\
$\eta$ & $(1,3,-1/3,1/3)$  & $(0,-1,0)^T$  & $(0,0,1)^T$ & $(+,+,-)^T$  \\
$\chi$&  $(1,3,-1/3,-2/3)$  & $(0,-1,0)^T$  & $(-1,-1,0)^T$ & $(-,-,+)^T$ \\
$\sigma$ & $(1,1,0,2)$ & $0$  & $2$ & $+$ \\
$\varphi$ & $(1,1,0,0)$ & $0$  & $0$ & $+$ \\
\hline
\end{tabular}
\caption{Summary of fermion and scalar fields, their representations under the gauge group 
\(\mathrm{SU}(3)_C \otimes \mathrm{SU}(3)_L \otimes \mathrm{U}(1)_X \otimes \mathrm{U}(1)_N\), 
electric charges (\(Q\)), \(B-L\) charges, and their transformations under the dark parity \(P_M\) symmetry.}
\label{table:1}
\end{table*}

Regarding gauge symmetries, this model shares most of its features with the rest of the 3-3-1 models. Specifically, the choice of the field representations under the \(\textrm{SU}(3)_{C} \otimes \textrm{SU}(3)_{L} \otimes \textrm{U}(1)_{X}\) symmetry is determined by anomaly cancellation, the allowance of tree-level mass generation, and ensuring that the SM fermions have the correct electric charges. It is also important to note that the \(\textrm{U}(1)_{X}\) charges are chosen such that no quarks possess exotic electric charges. The electric charge operator \(Q\) is defined as:
\begin{eqnarray}\label{eq:Q}
Q = T_{3} - \frac{1}{\sqrt{3}}T_{8} + X\,\mathbf{1}_{3\times3},
\end{eqnarray}
where \(T_{3}\) and \(T_{8}\) are the diagonal generators of \(\textrm{SU}(3)_{L}\).

Moreover, the representation of the fields under \(\textrm{U}(1)_{N}\) is chosen such that the SM fermions have the conventional \(B-L\) charges, and a discrete symmetry 
\begin{equation}
P_M = (-1)^{3(B-L)+2s},
\label{PMsymmetry}
\end{equation}
where \(s\) is the spin of the fields,
remains after spontaneous symmetry breaking \cite{Dimopoulos:1981zb,Farrar:1978xj,Ibanez:1991pr,Kadastik:2009cu,Kadastik:2009dj,Dong:2013wca,Alves:2016fqe,Dong:2014wsa,Dias:2022hbu}. Here, \(B-L\) is defined as:
\begin{eqnarray}\label{eq:BL}
B-L = -\frac{2}{\sqrt{3}}T_8 + N\,\mathbf{1}_{3\times 3},
\end{eqnarray}
 For example, in the particular case of the \(\textrm{U}(1)_{N}\) charge for the \(L_{iL}\) triplet, it should be chosen as \(-2/3\). A summary of all the fields and their representations under the gauge group is provided in Table~\ref{table:1}. 
 
 The \(P_M\) symmetry stabilizes several neutral fields that can, in principle, serve as candidates for dark matter. Finally, we note that a neutral fermion charged under the \(P_M\) symmetry, \(f_{iR} \sim (1,1,0,0)\), is introduced to also implement a see-saw mechanism in the DM leptonic sector.

\section{Scalar Sector}\label{scsect}
In this section, we study the scalar sector at tree level, as it plays a crucial role in understanding dynamical symmetry breaking. Additionally, it allows us to determine the mass eigenstates in different sectors, which we primarily classify as CP-even and CP-odd. In this model, the CP-odd sector is also $P_M$-odd. The $P_M$-odd sector refers to scalars that are non-trivially charged under the $P_M$ symmetry, which, in principle, stabilizes the dark matter sector. Due to our consideration that all quartic coupling constants  and vacuum expectation values are real, 
these sectors do not mix with each other. The charged scalar sector in this model is rather straightforward, 
as all charged fields, $(\eta_2^{-}, \chi_2^{-}, \eta_2^{+}, \chi_2^{+})$, 
exhibit a mass matrix that is identically zero. Consequently, these fields correspond to Nambu-Goldstone bosons, 
which are subsequently absorbed by the respective charged gauge fields.

\subsection{Scalar Potential}\label{sec:pot}

Let us consider the most general renormalizable, scale-invariant scalar potential that preserves the gauge symmetry of the model:
\begin{equation}
\begin{split}
V_{0} &= \lambda_{\chi}(\chi^{\dagger}\chi)^{2}+\lambda_{\eta} (\eta^{\dagger}\eta )^{2}+\lambda_{\eta\chi} (\chi^{\dagger}\chi ) (\eta^{\dagger}\eta)
\\
&+\lambda_{\eta\chi}^\prime (\chi^{\dagger}\eta ) (\eta^{\dagger}\chi )  + \lambda_{\chi\varphi} (\chi^{\dagger}\chi ) \varphi^2
\\
&+ \lambda_{\chi\sigma} (\chi^{\dagger}\chi )(\sigma^{*}\sigma ) + \lambda_{\eta\varphi} (\eta^{\dagger}\eta )\varphi^2 
\\
&+ \lambda_{\eta\sigma}  (\eta^{\dagger}\eta )  (\sigma^{*}\sigma ) +  \lambda_{\varphi}\varphi^4 + \lambda_{\sigma} (\sigma^{*}\sigma )^{2} 
\\
&+\lambda_{\varphi\sigma} (\sigma^{*}\sigma )\varphi^2.
\end{split}
\end{equation}

Since the potential \( V_0\) is biquadratic in the fields, it can be conveniently expressed as:
\begin{equation}
V_0 = (\Phi^{\circ 2})^T \Lambda(\theta) \Phi^{\circ 2},
\label{eq:V0:Hadamard}
\end{equation}
where the vector of field norms $(\Phi^{\circ 2})^T \equiv (|\eta|^2,|\chi|^2,|\sigma|^2,|\varphi|^2) \geq 0$, the symmetric matrix 
\begin{equation}
    \Lambda(\theta) = 
\begin{pmatrix}
  \lambda_{\eta} & \frac{1}{2} (\lambda_{\eta \chi } +  \lambda'_{\eta \chi } \theta) & \frac{1}{2} \lambda_{\eta \sigma}  & \frac{1}{2} \lambda_{\eta \varphi}  \\
\star &  \lambda_\chi & \frac{1}{2} \lambda_{\chi\sigma}  & \frac{1}{2} \lambda_{\chi \varphi} \\
\star & \star & \lambda_\sigma & \frac{1 }{2} \lambda_{\varphi \sigma} \\
\star & \star & \star &  \lambda_\varphi \\
\end{pmatrix}
\end{equation}
contains quartic couplings, and the orbit parameter $\theta \equiv (\chi^\dagger \eta) (\eta^\dagger \chi)/(|\eta|^2 |\chi|^2) \in [0,1]$ due to the Cauchy–Schwartz inequality. Notice here that we are using the Hadamard product, defined as the element-wise product, e.g. $[(A)^{\circ 2}]_{ij} = a_{ij}^2$ for a matrix $A$ with elements $a_{ij}$.

\subsection{Potential Minimum and $P_M$-even Mass Matrix}\label{sec:min}

We proceed to calculate the mass eigenvalues and eigenstates of the CP-even sector. To do this, we must first find the minimum of the scalar potential. Following standard procedures, we expand $V_0$ in terms of the fields $(\Phi^{\circ 2})^T = \frac{1}{2} (S_{\eta 1}^2, S_{\chi 3}^2, S_\sigma^2, S_\varphi^2)$, which acquire VEVs, while setting all other fields to zero. Additionally, we work within the Gildener-Weinberg approximation~\cite{Gildener:1976ih}, which allows us to calculate the effective potential perturbatively along the flat direction, as this method is convenient for models with multiple scalar fields.
Moreover, as we will see in Eq.~\eqref{lambdachieta}, $\lambda'_{\eta \chi }$ is positive, ensuring that, at the minimum of the potential, the orbit parameter $\theta$ equals zero. In other words, at the minimum
$V_0(\Phi) = (\Phi^{\circ 2})^T \Lambda (\theta = 0) \Phi^{\circ 2} \equiv (\Phi^{\circ 2})^T \Lambda \Phi^{\circ 2}$, where, for simplicity, we denote $\Lambda \equiv \Lambda(0)$.

To explicitly determine the flat direction, described by $\Phi = \xi \mathbf{n}$, where $\xi$ represents the radial degree of freedom (the dilaton) and $n$ is a unit vector, we proceed following~\cite{Kannike:2019upf}. The central idea is to minimise the potential $V_0(\Phi) = (\Phi^{\circ 2})^T \Lambda \Phi^{\circ 2}$, subject to the condition $V_0(\mathbf{n}) = 0$. As shown in detail in Ref.~\cite{Kannike:2019upf}, the Hadamard square of the unit vector $\mathbf{n}$ is an eigenvector of the quartic coupling matrix $\Lambda$ with a null eigenvalue, given by 
\begin{equation}
    \mathbf{n}^{\circ 2} = \frac{\adj(\Lambda) e}{e^T \adj(\Lambda) e}.
    \label{eq:n:solution}
\end{equation}
Here, $\adj (\Lambda)$ is the adjugate matrix of $\Lambda$, defined as $ \Lambda \, \adj (\Lambda) = \det(\Lambda) I,$ the vector $e = (1,\dots,1)^T$ on the respective space has all elements equal to unity, and $I$ is the identity matrix. The condition $V_0(\mathbf{n}) = 0$ is guaranteed by taking $\det (\Lambda) = 0$, as seen by inserting Eq.~\eqref{eq:n:solution} in Eq.~\eqref{eq:V0:Hadamard}.

The 1-loop effective potential along the flat direction is given by (see \cite{Coleman:1973jx,AlexanderNunneley:2010nw,Kannike:2019upf,Dias:2022hbu})
\begin{equation}\label{1-loop potential}
V^{1-\mathrm{loop}}(\xi \mathbf{n}) =  B(\mathbf{n}) \xi^4 \left( \ln \frac{\xi^2}{v_\xi^2} - \frac{1}{2} \right),
\end{equation}
where $\xi$ is the dilaton field, $v_\xi\equiv \sqrt{v_\eta^2 + v_\chi^2 + v_\sigma^2 + v_\varphi^2}$ its VEV, and
\begin{equation} \label{eq:Bn}
    B(\mathbf{n}) = \frac{1}{64\pi^2 v_\xi^4} \left( \sum_{S} m_S^4 + 3  \sum_V m_V^4- 4 \sum_F m_F^4 \right)
\end{equation}
is the sum over all degrees of freedom.

The mass matrix for the CP-even fields that get VEVs can be written as \footnote{Note that due to  normalisation of the fields, the numerical coefficient on the right-hand side is different from Ref.~\cite{Kannike:2019upf}.}
\begin{equation}
\label{Hessian matrix}
   M^2_\Phi = v_\xi^2 \; \nabla^T_\mathbf{N} \nabla_\mathbf{N} V(\mathbf{N})|_{\mathbf{N}=\mathbf{n}} = 2 \Lambda \circ (\mathbf{n} \mathbf{n}^T) \, v_\xi^2,
\end{equation}
where $\mathbf{N}$ is a unit vector in the $\Phi$ direction. Solving Eq.~\eqref{Hessian matrix} for the quartic coupling matrix $\Lambda$, we get
\begin{equation}\label{lambda phi}
\Lambda = \frac{1}{2 v_\xi^2} M_{\Phi}^2 \circ (\mathbf{n} \mathbf{n}^T)^{\circ -1},
\end{equation}
where $(\mathbf{n} \mathbf{n}^T)^{\circ -1}$ means the element-wise inverse, i.e. $[(\mathbf{n} \mathbf{n}^T)^{\circ -1}]_{ij} = 1/(n_i n_j)$ in Hadamard notation. 

Eq.~\eqref{lambda phi} is particularly useful as it enables us to express the quartic couplings in terms of the tree-level scalar masses, the VEVs, and mixing angles. To illustrate this, let us define the diagonal mass matrix $M_\textrm{CP-even}^2$ for the CP-even fields that acquire VEVs as $M_\textrm{CP-even}^2 = \textrm{diag}(m_{H_1}^2,\, m_{H_2}^2,\, m_{H_3}^2,\, m_{H_4}^2)$, which is related to $M_{\Phi}^2$ by   
\begin{equation}\label{relation mphi and md}
    M_{\Phi}^2 = O M_\textrm{CP-even}^2 O^{T},
\end{equation}
where the orthogonal matrix $O$ is given by \cite{PhysRevD.35.1732,SinghKoranga:2021oqf}
\begin{equation}\label{O matrix}
    O = O_{34} O_{24} O_{14} O_{23}O_{13}O_{12}.
\end{equation}
Each $O_{ij}$ defines a rotation in the $ij$-plane; we denote the mixing angle in $O_{ij}$ as $\alpha_{ij}$ and also abbreviate $s_{ij} \equiv \sin \alpha_{ij}$ and $c_{ij} \equiv \cos \alpha_{ij}$.

Considering $m_{H_4} = m_\xi = 0$ as the dilaton mass at tree level, 
we have the last column of $O$ as the flat direction:
\begin{equation}
\mathbf{n} = \begin{pmatrix} s_{14} & c_{14}s_{24} & c_{14}c_{24}s_{34} & c_{14}c_{24}c_{34} \end{pmatrix}^T. \label{eq:n:sol}
\end{equation}
Thus, the quartic couplings can be expressed in terms of the VEVs, CP-even scalar masses, and mixing angles using Eq.~\eqref{lambda phi}. Here, we note that the angles in Eq.~\eqref{eq:n:sol}, $\alpha_{14}, \alpha_{24}, \alpha_{34}$, can be easily written in terms of the VEVs as
\begin{equation} \label{alphas_123}
    \begin{split}
    %   v_\xi &= \sqrt{v_\eta^2 + v_\chi^2 + v_\sigma^2 + v_\varphi^2}, \\
        \alpha _{14} &= \arctan \left(\frac{v_\eta}{\sqrt{v_\chi^2 + v_\sigma^2+v_\varphi^2}}\right) ,\\
        \alpha_{24} &= \arctan \left(\frac{v_\chi}{\sqrt{v_\sigma^2+v_\varphi^2}}\right),\\ 
        \alpha_{34} &=\arctan \left(\frac{v_\sigma}{v_\varphi}\right). 
    \end{split}
\end{equation}
The remaining angles in Eq.~\eqref{O matrix}, $\alpha_{12}, \alpha_{13}$, and $\alpha_{23}$, cannot be expressed only in terms of the VEVs. However, they are strongly constrained by the perturbative unitarity conditions and the hierarchy in the VEVs, $v_\eta \ll v_\chi \ll v_\sigma, v_\varphi$ arising from the imposition of the see-saw mechanism throughout the fermion sector. In this model, it is natural to identify $v_\eta$ with the Standard Model VEV, $v_\textrm{SM} = 246.22$~GeV, since the $W^{\pm}$ mass, as shown in Eq.~\eqref{wv_masses}, is entirely determined by $v_\eta$. Thus, we identify the mass eigenstate $H_1$ as the SM-like Higgs with mass $m_{H_1} = 125.11$~GeV. The $H_2$ eigenstate acquires a mass at the $331$-breaking scale, $v_\chi$. Finally, the $H_3$ eigenstate acquires a mass at the scale of $v_\sigma$ and $v_\phi$, leading to the hierarchy $m_{H_1} \ll m_{H_2} \ll m_{H_3}$. The dilaton $H_4$ obtains a mass at one-loop level: it is given by
\begin{equation}
    m_{H_4}^2 = 8 B(\textbf{n}) v_\xi^2.
\end{equation}

For the SM-like Higgs $H_1 \approx S_{\eta 1}$, it is required that $\alpha_{12} \approx 0$. Additionally, to keep the couplings within the perturbative regime, the angles $\alpha_{13}$ and $\alpha_{23}$ must be small. To derive useful expressions for the CP-even eigenstates, we set these angles to zero. Under this approximation, we have:
{\footnotesize
\begin{equation}
    \begin{split}
         H_1 &= \frac{(v_\chi^2+v_\sigma^2 + v_\varphi^2) S_{\eta 1} - v_\eta (v_\chi S_{\chi 3} + v_\varphi S_\varphi)-v_\eta v_\sigma S_\sigma}{\sqrt{(v_\sigma^2 + v_\varphi^2) (v_\eta^2 + v_\sigma^2 + v_\varphi^2)}}  \approx S_{\eta 1}, \\
        H_2 &= \frac{(v_\sigma^2 + v_\varphi^2) S_{\chi 3} - v_\chi (v_\sigma S_\sigma + v_\varphi S_\varphi)}{\sqrt{(v_\eta^2 + v_\sigma^2 + v_\varphi^2) (v_\eta^2 + v_\chi^2 + v_\sigma^2 + v_\varphi^2)}}
        \approx S_{\chi 3}, \\
        H_3 &= \frac{v_\varphi S_\sigma - v_\sigma S_\varphi}{\sqrt{(v_\sigma^2 + v_\varphi^2)}}, \\
        H_4 &= \frac{v_\eta S_{\eta 1} + v_\chi S_{\chi 3} + v_\sigma S_\sigma + v_\varphi S_\varphi}{\sqrt{(v_\eta^2 + v_\chi^2 + v_\sigma^2 + v_\varphi^2)}} \approx 
        \frac{v_\sigma S_\sigma + v_\varphi S_\varphi}{\sqrt{(v_\sigma^2 + v_\varphi^2)}}.
    \end{split}
\end{equation}
}
\normalsize

Actually, we need to have small non-zero values for the angles $\alpha_{12}$, $\alpha_{13}$ and $\alpha_{23}$ to allow non-zero couplings of the fermion DM candidate to scalar field in order to ensure the correct relic density. 

\subsection{$P_M$-odd Sector}\label{sec:odd}

Finally, we focus our attention on the calculation of the $P_M$ scalar sector. In the basis, $B_d=(\eta_3, \chi_1^*)^T$, the $P_M$-odd mass matrix is
\begin{equation}\label{MPMO matrix}
   M_{d}^2 = 
   \begin{pmatrix}
 \frac{1}{2} \lambda'_{\eta \chi} v_\chi^2 & \frac{1}{2} \lambda'_{\eta \chi } v_\eta v_\chi \\
 \frac{1}{2} \lambda'_{\eta \chi } v_\eta v_\chi & \frac{1}{2} \lambda'_{\eta \chi } v_\eta^2 \\
\end{pmatrix}.
\end{equation}
The matrix $M_{d}$ is diagonalized by the orthogonal transformation $O_{d}^T M_d^2 O_d=\textrm{diag}(m_{G_d},m_{S_d})$, where
\begin{equation}
   O_{d}= \begin{pmatrix}
   -\frac{v_\eta}{\sqrt{v_\chi^2 + v_\eta^2}} &  \frac{v_\chi}{\sqrt{v_\chi^2 + v_\eta^2}} \\
   \frac{v_\chi}{\sqrt{v_\chi^2 + v_\eta^2}}  &    \frac{v_\eta}{\sqrt{v_\chi^2 + v_\eta^2}}    
    \end{pmatrix}.
\end{equation}
Its eigenvalues are $m^2_{G_d} = 0$ for a dark Goldstone and $m^2_{S_d} = \frac{1}{2} (v_\eta^2 + v_\chi^2) \lambda'_{\eta \chi }$ for a physical dark mass eigenstate $S_d$. We can now find the $ \lambda'_{\eta \chi }$ parameter:
\begin{equation}\label{lambdachieta}
   \lambda'_{\eta \chi } =  \frac{2m_{S_{d}}^2}{\left(v_\eta^2+ v_\chi^2\right)}.
\end{equation}
The dark scalar mass eigenstates are given by
\begin{equation}
\begin{split}
G_{d} &=  -\frac{v_\eta}{\sqrt{v_\chi^2 + v_\eta^2}} \eta_3  + \frac{v_\chi}{\sqrt{v_\chi^2 + v_\eta^2}} \chi_1^*,   \\
S_{d} &= \frac{v_\chi}{\sqrt{v_\chi^2 + v_\eta^2}} \eta_3  + \frac{v_\eta}{\sqrt{v_\chi^2 + v_\eta^2}} \chi_1^*. \label{Sd_scalar}
\end{split}
\end{equation}
Note that the scalar $S_d$ can, in principle, act as a dark matter candidate.

\section{Perturbative unitarity}\label{sec:pert}

Perturbative unitarity constraints arise from the requirement that the scattering matrix be unitary, ensuring the conservation of probability in particle interactions, and are explored by analyzing the tree-level two-to-two scattering amplitudes dominated by quartic interactions in the high-energy limit $\sqrt{s} \to \infty$. In this context, the eigenvalues $\mathcal{T}^{ba}$ of the scattering matrix must satisfy the condition $\Re\big(\, \mathcal{T}^{ba}_{0} \big)\leq \frac{1}{2}$ for the theory to remain perturbatively unitary.

At the zeroth partial wave order, in the high-energy limit, the matrix element can be expressed as \cite{Kannike_2024}:
\begin{equation}\label{matrix:T0}
    \mathcal{T}_0^{ba} = \frac{1}{16\pi} 
    \frac{1}{\sqrt{2^{\delta_{S_1,S_2}} 2^{\delta_{S_3,S_4}}}} 
    \frac{\partial V_0}{\partial S_1 \partial S_2 \partial S_3^* \partial S_4^*} \bigg|_{n} \leq \frac{1}{2},
\end{equation}
where \( a \) corresponds to the initial state \( S_1 + S_2 \) and \( b \) to the analogous final state \( S_3 + S_4 \) for the two-to-two scattering process.

By calculating the scattering matrix for all possible two-to-two initial and final states, including Goldstone bosons, we derive constraints on the quartic couplings. These constraints are given by:
{\footnotesize
\begin{equation}
    \begin{split}
        &|\lambda_\eta| \leq 4\pi, \quad |\lambda_{\eta \sigma}| \leq 8\pi, \quad |\lambda_{\eta \varphi}| \leq 8\pi, \quad |\lambda_\sigma| \leq 4\pi, \quad |\lambda_\chi| \leq 4\pi, \\
        &|\lambda_\eta - \sqrt{\lambda_\eta^2 - 2 \lambda_\eta \lambda_\chi + \lambda_{\eta \chi}^{'2} + \lambda_\chi^2} + \lambda_\chi| \leq 8\pi, \\
        &|\lambda_\eta + \sqrt{\lambda_\eta^2 - 2 \lambda_\eta \lambda_\chi + \lambda_{\eta \chi}^{'2} + \lambda_\chi^2} + \lambda_\chi| \leq 8\pi, \\
        &|\lambda_{\eta \chi} - \lambda'_{\eta \chi}| \leq 8\pi, \quad |\lambda_{\chi \sigma}| \leq 8\pi, \quad |\lambda_{\chi \varphi}| \leq 8\pi, \\
        &|\lambda_{\eta \chi}| \leq 8\pi, \quad |\lambda'_{\eta \chi} + \lambda_{\eta \chi}| \leq 8\pi, \quad |3 \lambda'_{\eta \chi} + \lambda_{\eta \chi}| \leq 8\pi.
    \end{split}
\end{equation}}
\normalsize
Additionally,  the four solutions $\lambda_i$  to the following quartic equation must satisfy $|\lambda_i| \leq 4\pi$:
\begin{equation}
    \lambda^4 + \lambda^3 A + \lambda^2 B + \lambda C + D = 0,
\end{equation}
with coefficients explicitly given as
\begin{equation}
    \begin{aligned}
        A &= -\frac{1}{2}  (8 \lambda_\eta +4 \lambda_\sigma +3 \lambda_\varphi +8 \lambda_\chi ),
    \end{aligned}
\end{equation}
\begin{equation}
    \begin{aligned}
        B &= \frac{1}{8} (-64 \lambda_\eta  \lambda_\sigma -48 \lambda_\eta  \lambda_\varphi -128 \lambda_\eta  \lambda_\chi +6 \lambda_{\eta \sigma}^2
        \\
        &+3 \lambda_{\eta \varphi} ^2+2 \lambda_{\chi \eta}^{\prime 2}+12 \lambda'_{\chi \eta} \lambda_{\chi \eta} -24 \lambda_\sigma  \lambda_\varphi -64 \lambda_\sigma  \lambda_\chi\\
        & -48 \lambda_\varphi  \lambda_\chi +\lambda_{\varphi \sigma }^2+18 \lambda_{\chi \eta} ^2+6 \lambda_{\chi \sigma}^2+3 \lambda_{\chi \varphi} ^2 ),
    \end{aligned}
\end{equation}
\begin{equation}
    \begin{aligned}
        C &= \frac{1}{8} (9 \lambda_{\eta \sigma}^2 \lambda_\varphi - 
   96 \lambda_\eta \lambda_\sigma \lambda_\varphi + \lambda_{\chi\eta}^{\prime 2} (4 \lambda_\sigma + 
      3 \lambda_\varphi)
      \\
      &+ 
   4 \lambda_\eta \lambda_{\varphi \sigma}^2 + 
   24 \lambda_{\eta \sigma}^2 \lambda_\chi - 
   256 \lambda_\eta \lambda_\sigma \lambda_\chi  \\
   &- 
   192 \lambda_\eta \lambda_\varphi \lambda_\chi - 
   96 \lambda_\sigma \lambda_\varphi \lambda_\chi + 
   4 \lambda_{\varphi \sigma}^2 \lambda_\chi 
      \\
      &+ 
   6 \lambda_{\eta \varphi}^2 (\lambda_\sigma + 
      2 \lambda_\chi) + 
   36 \lambda_\sigma \lambda_{\chi \eta}^2 + 
   27 \lambda_\varphi \lambda_{\chi \eta}^2  \\
   &- 
   18 \lambda_{\eta \sigma} \lambda_{\chi \eta} \lambda_{\chi \sigma} + 24 \lambda_\eta \lambda_{\chi \sigma}^2 + 
   9 \lambda_\varphi \lambda_{\chi \sigma}^2 
   \\
   & - 
   3 \lambda_{\varphi \sigma} \lambda_{\chi \sigma} \lambda_{\chi \varphi} + 12 \lambda_\eta \lambda_{\chi \varphi}^2 + 
   6 \lambda_\sigma \lambda_{\chi \varphi}^2  \\
   &+ 
   3 \lambda_{\chi \eta}' (8 \lambda_\sigma \lambda_{\chi \eta} + 
      6 \lambda_\varphi \lambda_{\chi \eta} - 
      2 \lambda_{\eta \sigma} \lambda_{\chi \sigma}  
      \\
      &- \lambda_{\eta \varphi} \lambda_{\chi \varphi}) - 
   3 \lambda_{\eta \varphi} (\lambda_{\eta \sigma} \lambda_{\varphi \sigma} + 
      3 \lambda_{\chi \eta} \lambda_{\chi \varphi})) ,
        \end{aligned}
\end{equation}
\begin{equation}
    \begin{aligned}
        D &= \frac{1}{32}  (\lambda_{\chi\eta}^{\prime 2} (-24 \lambda_\sigma \lambda_\varphi + \lambda_{\varphi \sigma}^2) - 
    96 \lambda_{\eta \varphi}^2 \lambda_\sigma \lambda_\chi 
    \\
    & - 144 \lambda_{\eta \sigma}^2 \lambda_\varphi \lambda_\chi 
+ 1536 \lambda_\eta \lambda_\sigma \lambda_\varphi \lambda_\chi \\
&+ 
    48 \lambda_{\eta \sigma} \lambda_{\eta \varphi} \lambda_{\varphi \sigma} \lambda_\chi - 
    64 \lambda_\eta \lambda_{\varphi \sigma}^2 \lambda_\chi - 216 \lambda_\sigma \lambda_\varphi \lambda_{\chi \eta}^2 
    \\
&+ 9 \lambda_{\varphi \sigma}^2 \lambda_{\chi \eta}^2 + 
    108 \lambda_{\eta \sigma} \lambda_\varphi \lambda_{\chi \eta} \lambda_{\chi \sigma} 
    \\
    &- 
    18 \lambda_{\eta \varphi} \lambda_{\varphi \sigma} 
\lambda_{\chi \eta} \lambda_{\chi \sigma}
+ 
    9 \lambda_{\eta \varphi}^2 \lambda_{\chi \sigma}^2 - 
    144 \lambda_\eta \lambda_\varphi
\lambda_{\chi \sigma}^2\\
&+ 
    72 \lambda_{\eta \varphi} \lambda_\sigma \lambda_{\chi \eta} \lambda_{\chi \varphi} - 
    18 \lambda_{\eta \sigma} \lambda_{\varphi \sigma} \lambda_{\chi \eta} \lambda_{\chi \varphi}  \\
&- 
    18 \lambda_{\eta \sigma} \lambda_{\eta \varphi} \lambda_{\chi \sigma} \lambda_{\chi \varphi} + 
    48 \lambda_\eta \lambda_{\varphi \sigma} \lambda_{\chi \sigma} \lambda_{\chi \varphi} 
    \\
    &+ 
    9 \lambda_{\eta \sigma}^2 \lambda_{\chi \varphi}^2 - 
    96 \lambda_\eta \lambda_\sigma \lambda_{\chi \varphi}^2 
- 6 \lambda_{\chi \eta}'  (24 \lambda_\sigma
\lambda_\varphi \lambda_{\chi \eta} 
\\
&- \lambda_{\varphi \sigma}^2 \lambda_{\chi \eta} - 
       6 \lambda_{\eta \sigma} \lambda_\varphi \lambda_{\chi \sigma} 
       + \lambda_{\eta \varphi} \lambda_{\varphi \sigma} 
\lambda_{\chi \sigma}\\
& - 4 \lambda_{\eta \varphi} \lambda_\sigma \lambda_{\chi \varphi} + \lambda_{\eta \sigma} \lambda_{\varphi \sigma} 
\lambda_{\chi \varphi})).
    \end{aligned}
\end{equation}

\section{Gauge Sector}\label{sec:gauge}

The gauge sector spectrum arises from the kinetic terms of the scalar bosons, which are given by 
\begin{equation}\label{eq:Lscalars_1}
    \mathcal{L} \supset (D_\mu \eta)^\dagger (D^\mu \eta) + (D_\mu \chi)^\dagger (D^\mu \chi) + (D_\mu \sigma)^* (D^\mu \sigma),
\end{equation}
where the covariant derivative for triplets is defined as 
\begin{equation}\label{covariant_derivative_3311}
    D_\mu = \partial_\mu - i g W_\mu^a T_a - i g_X X B_\mu - i g_N N N_\mu.
\end{equation}
Here, \( T_a = \frac{\lambda_a}{2} \) are the generators of $\textrm{SU}(3)_{L}$, with \( \lambda_a \) (\( a = 1, \dots, 8 \)) being the Gell-Mann matrices. The parameters \( X \) and \( N \) represent the charges of the fields under the $\textrm{U}(1)_{X}$ and $\textrm{U}(1)_{N}$ symmetries, respectively. Note that the \( \varphi \) field is a gauge singlet and, therefore, its kinetic term is not included in Eq.~\eqref{eq:Lscalars_1}. The covariant derivative for the \( \sigma \) field only involves the \( N_\mu \) gauge boson, as \( \sigma \) carries a charge of two under the $\textrm{U}(1)_{N}$ symmetry.

Since \( \textrm{SU}(2)_{L} \subset \textrm{SU}(3)_{L} \), the gauge coupling \( g \) is the same as in the SM. The gauge bosons associated with the \( \textrm{SU}(3)_{L} \), \( \textrm{U}(1)_{X} \), and \( \textrm{U}(1)_{N} \) symmetries are \( W_\mu^a \), \( B_\mu \), and \( N_\mu \), respectively, with their corresponding gauge couplings denoted as \( g \), \( g_X \), and \( g_N \).

The charged (including the non-Hermitian) gauge bosons are straightforwardly calculated, as they arise from the non-diagonal generators of the $\textrm{SU}(3)_L$ group. The charged gauge eigenstates are given by
\begin{align}
W^\pm_\mu &= \frac{W_{1\mu} \mp i W_{2\mu}}{\sqrt{2}}, \notag \\ 
V^{0(\dagger)}_\mu &= \frac{W_{4\mu} \mp i W_{5\mu}}{\sqrt{2}}, \label{charged_gauge_bosons} \\ 
V^\pm_\mu &= \frac{W_{6\mu} \pm i W_{7\mu}}{\sqrt{2}}\,, \notag 
\end{align}
with masses
\begin{equation}\label{wv_masses}
\begin{aligned}
    m_{W^\pm}^2 &= \frac{g^2 v_\eta^2}{4} \,, \\
    m_{V^{0(\dagger)}}^2 &= \frac{g^2}{4} \left(v_\eta^2 + v_\chi^2\right) \,, \\
    m^2_{V^\pm} &= \frac{g^2 v_\chi^2}{4}\,.
\end{aligned}
\end{equation}
Note that the fields \( V^\pm_\mu \) and \( V^{0(\dagger)}_\mu \) are \( P_M \)-odd, or dark, under the \( P_M \) symmetry defined in Eq.~\eqref{PMsymmetry}. However, since the neutral gauge boson is heavier than the charged one, i.e., \( m_{V^{0(\dagger)}}^2 = m_{W^\pm}^2 + m^2_{V^\pm} \), the dark gauge sector does not provide a suitable DM candidate.

On the other hand, diagonalization of the squared mass matrix of the neutral gauge bosons is more intricate. In the basis 
\[
{\bf B}_{\rm NG} = (W_{3\mu}, W_{8\mu}, B_\mu, N_\mu)\,,
\]
the mass matrix is given by
\begin{widetext}
\begin{equation}\label{NG_mass_matrix}
    M_{\mathrm{NG}}^2 =
    \frac{g^2}{4}\begin{pmatrix}
        v_\eta^2 & \frac{v_\eta^2}{\sqrt{3}} & -\frac{2g_X v_\eta^2}{3 g} & \frac{2g_N v_\eta^2}{3 g} \\
        \frac{v_\eta^2}{\sqrt{3}} & \frac{v_\eta^2+4 v_\chi^2}{3} & -\frac{2g_X \left(v_\eta^2-2 v_\chi^2\right)}{3\sqrt{3}g} & \frac{2 g_N \left(v_\eta^2+4 v_\chi^2\right)}{3 \sqrt{3}g} \\
        -\frac{2 g_X v_\eta^2}{3g} & -\frac{2g_X \left(v_\eta^2-2 v_\chi^2\right)}{3\sqrt{3}g} & \frac{4g_X^2 \left(v_\eta^2+v_\chi^2\right)}{9g^2} & -\frac{4 g_N g_X \left(v_\eta^2-2 v_\chi^2\right) }{9g^2} \\
        \frac{2g_N v_\eta^2}{3 g} & \frac{2 g_N \left(v_\eta^2+4 v_\chi^2\right)}{3 \sqrt{3}g} & -\frac{4 g_N g_X \left(v_\eta^2-2 v_\chi^2\right) }{9g^2} & \frac{4 g_X^2 \left(v_\eta^2+4 \left(9 v_\sigma^2+v_\chi^2\right)\right)}{9g^2}
    \end{pmatrix}.
\end{equation}
\end{widetext}

For simplicity, we are neglecting the kinetic mixing term between the Abelian fields \( B_\mu \) and \( N_\mu \).

Since there is a hierarchy between the VEVs, at leading order, the mass matrix \( M_{\mathrm{NG}}^2 \) is diagonalized by the orthogonal transformation matrix \( U_{\mathrm{NG}} \), such that
\[
U_{\mathrm{NG}}^T M_{\mathrm{NG}}^2 U_{\mathrm{NG}} = \textrm{diag}(0, m_Z^2, m_{Z'}^2, m_{Z''}^2),
\]
where \( U_{\mathrm{NG}} \) is approximately

\begin{widetext}
\begin{equation}
U_{\mathrm{NG}} \simeq 
\begin{pmatrix}
    \sin\theta_W &
    \cos\theta_W & 
    -\frac{\sqrt{3} g \tan^2 \theta_W v_\eta^2}{2 g_X \tan 2\theta_W v_\chi^2} &
    \frac{g v_\eta^2}{24 g_N v_\sigma^2} \\
    -\frac{\sin\theta_W}{\sqrt{3}} &
    \frac{\sin^2 \theta_W }{\sqrt{3}\cos\theta_W} &
    -\frac{g \tan\theta_W}{g_X} &
    \frac{g v_\chi^2}{6\sqrt{3} g_N v_\sigma^2} \\
    \frac{g\sin\theta_W}{g_X} &
    -\frac{g \sin^2 \theta_W}{g_X \cos\theta_W} & 
    -\frac{\tan\theta_W}{\sqrt{3}} & 
    \frac{g_X v_\chi^2}{18 g_N v_\sigma^2} \\
    0 &
    -\frac{g \tan^2 \theta_W v_\eta^2}{24\cos\theta_W g_N v_\sigma^2} & 
    \frac{g_X v_\chi^2}{6\sqrt{3} g_N \tan\theta_W v_\sigma^2} &
    1
\end{pmatrix}.
\label{eq:UNG}
\end{equation}
\end{widetext}
Here, the Weinberg angle \( \theta_W \) is defined as
\begin{equation}
    \sin^2\theta_W = \frac{g_X^2}{\sqrt{g^2+\frac{4}{3}g_X^2}}.
\end{equation}
Since \( g \) can be approximately set to \( 0.652 \), based on the \( W^{\pm} \) mass, and using \(\sin^2\theta_W \approx 0.231\)~\cite{ParticleDataGroup:2024cfk}, the value of \( g_X \) is found to be approximately \( 0.377 \).

The \( U_{\mathrm{NG}} \) matrix allows expressing the gauge symmetry bosons, \( W_{3\,\mu}, W_{8\,\mu}, B_{\mu}, N_{\mu} \), in terms of the physical gauge bosons, \( A_\mu, Z_\mu, Z_\mu^{\prime}, Z_\mu^{\prime\prime} \). That is,
\begin{equation}
\begin{pmatrix}
    W_{3\,\mu} \\ W_{8\,\mu} \\ B_{\mu} \\ N_{\mu}    
\end{pmatrix}
= U_{\mathrm{NG}}
\begin{pmatrix}
    A_\mu \\ Z_\mu \\ Z_\mu^{\prime} \\ Z_\mu^{\prime\prime}    
\end{pmatrix}.
\end{equation}
Besides the fact that the photon mass is zero, we can obtain the masses of the \( Z \), \( Z' \), and \( Z'' \) bosons:
% \begin{widetext} 
% \begin{align}
%     m_{Z}^2 &\simeq  
%     \frac{9 g^2 v_\eta^2 v_\chi^2 \cos^2 \theta_W}{
%     \sin^4 \theta_W \left( 
%         v_\eta^2 \left( 36  + v_\chi^2 \right) + 36  v_\chi^2 
%     \right) 
%     - 36  \sin^2 \theta_W \left( v_\eta^2 + 2 v_\chi^2 \right)
%     + 9  \left( v_\eta^2 + 4 v_\chi^2 \right) } \,, \\
%     m_{Z^{\prime}}^2 &\simeq \frac{g_X^2 v_\chi^2}{3 \tan^2 \theta_W} \,, \\
%     m_{Z^{\prime\prime}}^2 &\simeq 4 g_N^2 v_\sigma^2 \,. \label{m_z2}
% \end{align}
% \end{widetext}
\begin{align}
       m_{Z}^2 &\simeq  
    [9 g^2 v_\eta^2 v_\chi^2 \cos^2 \theta_W]/[
    \sin^4 \theta_W ( 
        v_\eta^2 \left( 36  + v_\chi^2 \right) 
        \notag
        \\
        &+ 36  v_\chi^2 
    ) 
    - 36  \sin^2 \theta_W ( v_\eta^2 + 2 v_\chi^2 )
    \notag
    \\
    &+ 9  ( v_\eta^2 + 4 v_\chi^2 ) ]\, ,
    \\
    m_{Z^{\prime}}^2 &\simeq \frac{g_X^2 v_\chi^2}{3 \tan^2 \theta_W} \,, \\
    m_{Z^{\prime\prime}}^2 &\simeq 4 g_N^2 v_\sigma^2 \,. \label{m_z2}
\end{align}

The masses \( m_{Z^{\prime}} \) and \( m_{Z^{\prime\prime}} \) were computed at leading order, while \( m_Z \) was computed at the next order, where it deviates from its SM counterpart. However, at leading order, \( m_Z \) coincides with the SM \( Z \) boson mass, as expected. Additionally, the parameter \( g_N \), together with \( v_\sigma \), determines the mass of the \( Z'' \), as given in Eq.~\eqref{m_z2}. For \( v_\sigma \) on the order of \( 10^3 \, \textrm{TeV} \) and Yukawa couplings of order 1, the potential remains stable (i.e., \( B(\textbf{n}) > 0 \), as defined in Eq.~\eqref{eq:Bn}) if \( g_N \gtrsim 0.598 \).

The \( Z \) boson properties in our model deviate from the SM predictions due to the (small) mixing between \( Z \) and the other massive vector fields. For example, we can obtain a lower bound on \( v_\chi \) by imposing that the new contribution to the \( \rho_0 \) parameter, defined as \( \rho_0 \equiv \left(\frac{M_W}{M_Z \cos\theta_W}\right)^2 \), remain within the experimental limits~\cite{Workman:2022ynf}. This parameter, which is exactly equal to unity in the SM, is modified in our model as:
\begin{equation}
    \Delta \rho_0 = \rho_0 - 1 \approx \frac{v_\eta^2 \cos^2 2\theta_W \sec^4 \theta_W}{4 v_\chi^2} \,.
\end{equation}
Using \( \sin^2\theta_W \approx 0.231 \) and the allowed experimental deviation \( \Delta \rho_0 \lesssim 5.8 \times 10^{-4} \)~\cite{Workman:2022ynf}, we find that
\begin{equation}
v_\chi \gtrsim 3571.35 \, \textrm{GeV}, \label{eq:lower_bound_vchi}
\end{equation}
setting the lower bound for the symmetry breaking scale of the $331$ model.

\section{Yukawa Sector: see-saw-Suppressed Fermion Masses}\label{scYukawa}

In this section, we derive the fermion mass spectrum, demonstrating how see-saw mechanisms emerge across all sectors as a result of the VEV hierarchy $v_\eta \ll v_\chi \ll v_\varphi, v_\sigma$. The electroweak VEV, $v_\eta$, is fixed at $v_{\rm SM} = 246$ GeV, while the 3-3-1 breaking VEV, $v_\chi$, is significantly larger, approximately $3.6$ TeV. An even larger VEV, $v_\varphi$, is required to enable the see-saw mechanism for charged fermions, with $v_\varphi$ typically around $10^6$ GeV. Finally, the $B-L$ symmetry-breaking VEV, $v_\sigma$, must also be large, satisfying $v_\sigma \gg v_\eta$, to justify the see-saw mechanism for neutrinos. This hierarchical structure plays a crucial role in shaping the Yukawa sector.

These see-saw mechanisms not only provide a partial explanation for the flavour puzzle but also help identify the fermionic dark matter candidate in our model. By incorporating this hierarchy, the model successfully reproduces the observed fermion mass spectrum and offers a natural framework to address open questions in particle physics, such as the flavour puzzle and the nature of dark matter.
\vspace{-\baselineskip}

\subsection{Lepton Spectrum}
The renormalizable Yukawa Lagrangian for the lepton sector is expressed as:
\be
\begin{split} \label{eq:LepYuk}
-\mathcal{L}_{l} &= Y^\nu_{ij}\,\overline{L_{iL}}\, \eta\, \nu_{jR} + Y^f_{ij}\,\overline{L_{iL}}\, \chi\, f_{jR} 
\\
&+ \frac{h_{ij}^{\nu}}{2} \,\sigma\, \overline{(\nu_{iR})^c}\, \nu_{jR} 
+ \frac{h^{f}_{ij}}{2}\, \varphi\, \overline{(f_{iR})^c} \,f_{jR} \\
& + y^\mathcal{F}_{ij}\,\overline{L_{iL}}\, \eta^* \,\mathcal{F}_{jR} + y^e_{ij}\,\overline{\mathcal{F}_{iL}}\, \chi^* \,e_{jR} 
\\
&+ h^\mathcal{F}_{ij}\,\varphi\, \overline{\mathcal{F}_{iL}}\, \mathcal{F}_{jR} + \textrm{h.c.}\,
\end{split}
\ee

Following spontaneous symmetry breaking, all leptons acquire mass, as will be demonstrated in the subsequent analysis. It is also noteworthy that all Yukawa couplings in the lepton sector are represented by $3 \times 3$ matrices. For simplicity, we omit explicit indices and work directly with the couplings in matrix form. 
%Furthermore, we assume that all Yukawa parameters are real, as CP violation is not considered in this analysis.
% \blue{We assume all Yukawa couplings to be real as we want to avoid stringent constraints from lepton electric dipole moment. We are mainly interested in the dark matter phenomenology and complex phases would only make numerical scans more cumbersome, without providing any additional insight. }

Moreover, to ensure compatibility with stringent experimental constraints on CP violation, we assume all Yukawa matrices to be real at leading order. The current 90\% confidence level limits, $|d_e| < 4.1 \times 10^{-30}\,e\cdot\mathrm{cm}$ from the ACME III experiment~\cite{Roussy2023}, and $|d_n| < 1.8 \times 10^{-26}\,e\cdot\mathrm{cm}$ from neutron EDM searches~\cite{nEDM2020}, already exclude $\mathcal{O}(1)$ CP phases in many extensions of the Standard Model. In 3-3-1 frameworks, even a single physical phase in the scalar or Yukawa sector typically drives the predicted $d_{e,n}$ close to—or above—these bounds~\cite{DeConto2016, Okada2022, Doff2006}. Our choice aligns with the paradigm of \textit{additive CP violation}, where tiny ($\lesssim 10^{-3}$) imaginary perturbations can reproduce CKM/PMNS observables while naturally suppressing EDMs~\cite{Dent2002}. Since the central aim of this work is to present a classically scale-invariant $\mathrm{SU(3)}_C \otimes \mathrm{SU(3)}_L \otimes \mathrm{U(1)}_X \otimes \mathrm{U(1)}_N$ model that addresses the fermion mass hierarchy through a universal see-saw mechanism and provides a naturally fermion dark matter candidate, introducing general complex phases would enlarge the parameter space without altering these core results. A comprehensive study of CP-violating effects, including renormalization group running and a SMEFT analysis of dipole operators, will be presented elsewhere~\cite{SMEFT2021}.

\subsubsection{Charged Leptons}\label{sssec:cleptons}

The charged lepton sector consists of $P_M$-even fields ($e_i$ and $E'_i$), which mix to produce the SM charged leptons and three heavy charged leptons, as well as $P_M$-odd fields ($E_i$), where $i = 1, 2, 3$.

In the basis ${\bf E}_{L,R} = (e_{L,R},\,E'_{L,R})$, the mass terms are expressed as $\overline{{\bf E}_{L}} \mathcal{M}_{e} {\bf E}_{R}$, where the mass matrix $\mathcal{M}_{e}$ is given by:
\bea \label{eq:clmassM}
\mathcal{M}_{e} =
\begin{pmatrix}
   0 & -y^{\mathcal{F}}\, v_\eta \\
   y^{e}\, v_\chi & h^{\mathcal{F}}\, v_\varphi
\end{pmatrix}\,.
\eea
This mass matrix exhibits a type-I Dirac see-saw structure in the limit $v_\eta \ll v_\chi \ll v_\varphi$. To leading order, the matrix $\mathcal{M}_e$ can be block-diagonalized using two unitary matrices, $V_L^e$ and $V_R^e$. The diagonalization matrices are given by:
\begin{equation}
\begin{split}
    V_L^e &= \begin{pmatrix}
    1 & -\frac{v_\eta}{v_\varphi}y^{\mathcal{F}} (h^{\mathcal{F}})^{-1} \\
    \frac{v_\eta}{v_\varphi}(h^{\mathcal{F}\dagger})^{-1}y^{\mathcal{F}\dagger} & 1
    \end{pmatrix}, \\
    V_R^e &= \begin{pmatrix}
    1 & \frac{v_\chi}{v_\varphi}y^{e\dagger} (h^{\mathcal{F}\dagger})^{-1} \\
    \frac{v_\chi}{v_\varphi}(h^{\mathcal{F}})^{-1}y^{e} & 1
    \end{pmatrix},    
\end{split}
\end{equation}
such that  
\begin{equation}
    V_L^{e\dagger} \mathcal{M}_e V_R^e = \textrm{diag}(M_e, M_{E'}),
\end{equation}
where the mass matrices are given by
\begin{equation}\label{e and Ep mass matrices}
    M_e = \frac{v_\eta v_\chi}{\sqrt{2} v_\varphi} y^{\mathcal{F}}(h^{\mathcal{F}})^{-1}y^e, \quad 
    M_{E'} = \frac{v_\varphi}{\sqrt{2}} h^{\mathcal{F}}.
\end{equation}
The mass matrix of $P_M$-odd fields, $E_i$, is 
\be\label{E-lepton mass matrix}
M_E=\frac{v_\varphi}{\sqrt{2}}h^{\mathcal{F}},
\ee
that is they are essentially degenerate with $P_M$-even fields $E'_i$.

For simplicity, we assume $y^{\mathcal{F}} = h^{\mathcal{F}}$ equal to the identity matrix and $y^e$ to be diagonal. This allows us to express the matrix elements in terms of the masses of the charged leptons $e$, $\mu$, and $\tau$. The diagonal Yukawa coupling matrix is then given by:
\begin{equation}
y^e = \mathrm{diag}(y^e_1, y^e_2, y^e_3).
\end{equation}
By solving for the matrix elements, we find the following relations:
\begin{equation}
    y^e_1 = \frac{\sqrt{2}v_\varphi}{v_\eta v_\chi} m_e, \quad
    y^e_2 = \frac{\sqrt{2}v_\varphi}{v_\eta v_\chi} m_\mu, \quad
    y^e_3 = \frac{\sqrt{2}v_\varphi}{v_\eta v_\chi} m_\tau. \label{ye_matrix}
\end{equation}
Considering the masses given in \cite{Workman:2022ynf}, with $v_\eta = v_{\rm SM}$, $v_\chi = 5 \cdot  10^3$ GeV, and $v_\varphi = 10^6$ GeV, the Yukawa parameters $y^e$ are of the orders:
\begin{equation}
   y^e_1 \sim 10^{-4}, \quad y^e_2 \sim 10^{-2}, \quad y^e_3 \sim 1.
\end{equation}
\subsubsection{Neutrinos}

The neutral lepton spectrum can be divided into the $P_M$-even (visible) and $P_M$-odd (dark) sectors. The $P_M$-even sector corresponds to neutrinos. We define the basis ${\bf N}_L = (\nu_L, (\nu_R)^c)$ and write the mass term as $(1/2)\,\overline{{\bf N}_L} \,\mathcal{M}_\nu\, ({\bf N}_L)^c$, where the mass matrix $\mathcal{M}_\nu$ is given by:
\bea \label{eq:numassM}
\mathcal{M}_\nu 
= \frac{1}{\sqrt{2}}
\begin{pmatrix}
   0 & Y^{\nu}\,v_\eta \\
   Y^{\nu T}\,v_\eta & h^{\nu}\,v_\sigma
\end{pmatrix} \,.
\eea 
Since $v_\sigma \gg v_\eta$, the matrix $\mathcal{M}_\nu$ exhibits the structure of a type-I see-saw mechanism \cite{Grimus:2000vj}. To leading order, it can be block-diagonalized by the matrix $V^\nu$, given by:
\begin{equation}
V^\nu = 
\begin{pmatrix}
    1 & \frac{v_\eta}{v_\sigma} Y^{\nu\ast} (h^{\nu\ast})^{-1} \\
    -\frac{v_\eta}{v_\sigma} (h^{\nu T})^{-1} Y^{\nu T} & 1
\end{pmatrix},
\end{equation}
such that 
\begin{equation}
V^{\nu T} \mathcal{M}_\nu V^\nu = \textrm{diag}(m_\nu,\, m_{N}),
\end{equation}
where the mass matrices are:
\begin{equation}
m_\nu = -\frac{v_\eta^2}{\sqrt{2} v_\sigma} Y^\nu (h^\nu)^{-1} Y^{\nu T}, \quad 
m_{N} = \frac{v_\sigma}{\sqrt{2}} h^\nu.
\end{equation}
For simplicity, we assume $h^\nu$ to be diagonal, as sterile neutrino oscillations are not considered in this work. Using the Casas-Ibarra parametrisation (\cite{Casas_2001}, \cite{Karkkainen:2021tbh}), and considering the $R$ matrix as the identity matrix, we obtain the solution for $Y^\nu$ 
\begin{equation}
\begin{split}
    Y^\nu &= -i \frac{\sqrt{\sqrt{2} v_\sigma}}{v_\eta} (m_\nu^{\rm diag})^{1/2} U_{2\nu}^{-1}\\ 
    &= -i \frac{\sqrt{\sqrt{2} v_\sigma}}{v_\eta} (m_\nu^{\rm diag})^{1/2} (U_{\rm PMNS})^T. \label{Y_nu_matrix}
\end{split}
\end{equation}
In the last equality, we use \( U_{2\nu} = U_{\rm PMNS}^*\) \cite{esteban2024nufit60updatedglobalanalysis}, as the charged lepton mass matrix is diagonal under our assumptions. Hence, this result expresses the elements of the Yukawa matrix \( Y^\nu \) in terms of the neutrino masses \( m_{\nu i} \) (for \( i=1,2,3 \)), which correspond to the physical neutrinos \( \nu_{iL} \), and the neutrino mixing angles in \( U_{\rm PMNS} \).

\subsubsection{Dark Leptons}

The $P_M$-odd, or dark, neutral leptons are arranged in the basis ${\bf F}_L=(f_{L},\, (f_{R})^c,\, (f^\prime_{L}), \, (f^\prime_{R})^c)$, such that their mass matrix, $(1/2)\,\overline{{\bf F}_{L}}\, \mathcal{M}_f\, ({\bf F}_{L})^c$, is expressed as
\bea \label{eq:darknumassM}
\mathcal{M}_{f}=
\begin{pmatrix}
   0_{[3\times 3]} & m^f_{D\,[3\times 9]}\\
   (m_{D}^{f})^T_{[9\times 3]}  & m^f_{R\,[9\times 9]}
\end{pmatrix},
\eea 
where 
\be
\begin{split}
m_D^f &= \frac{1}{\sqrt{2}}
\begin{pmatrix}
    Y^{f}\,v_\chi & 0 & -y^{\mathcal{F}}\,v_\eta
\end{pmatrix}, \\
m_R^f &= \frac{1}{\sqrt{2}}
\begin{pmatrix}
    h^{f}\,v_\varphi & 0 & 0 \\
    0 & 0 & h^{\mathcal{F}}\,v_\varphi  \\
    0 &  h^{\mathcal{F}\,T}\,v_\varphi & 0
\end{pmatrix}. 
\end{split}
\ee
The structure corresponds to a type-I see-saw, as the eigenvalues of $m^f_R$ are proportional to $v_\varphi$, while the entries of $m^f_D$ are at most proportional to $v_\chi$, with $v_\varphi \gg v_\chi \gg v_\eta$. Therefore, the same procedure applied to neutrinos can be repeated. In this case, however, block diagonalization requires two matrices, $V_1^f$ and $V_2^f$. The transformation is given by
\begin{equation}
V_2^{f T} V_1^{f T} \mathcal{M}_f V_1^f V_2^f = \textrm{diag}(m_f,\, m_{f'_1},\, m_{f'_2},\, m_{f'_3}), \nonumber
\end{equation}
where the block matrices on the diagonal are
\begin{align}
\begin{aligned}
m_f &= -\frac{v_\chi^2}{\sqrt{2} v_\varphi} Y^f (h^{\mathcal{F}})^{-1} Y^{f T}, \\
m_{f'_1} &= \frac{v_\varphi}{\sqrt{2}} h^f, \\
m_{f'_2} &= \frac{v_\varphi}{\sqrt{2}} h^{\mathcal{F}}, \\
m_{f'_3} &= -\frac{v_\varphi}{\sqrt{2}} h^{\mathcal{F}}. \label{dark_leptons}
\end{aligned}
\end{align}
The matrices $V_2^f$ and $V_1^f$ are explicitly given by:
\begin{equation}
\begin{aligned}
V_2^f &= \begin{pmatrix}
    1 & 0 & 0 & 0 \\
    0 & 1 & 0 & 0 \\
    0 & 0 & \frac{1}{\sqrt{2}} & -\frac{1}{\sqrt{2}} \\
    0 & 0 & \frac{1}{\sqrt{2}} & \frac{1}{\sqrt{2}}
\end{pmatrix}.
\end{aligned}
\end{equation}
{\footnotesize
\begin{equation}
\begin{aligned}
V_1^f &= \begin{pmatrix}
    1 & 
    \frac{v_\chi}{v_\varphi} Y^{f\ast} (h^{f\ast})^{-1} & 
    -\frac{v_\eta}{v_\varphi} y^{\mathcal{F}\ast} (h^{\mathcal{F}\ast})^{-1} & 
    0 \\
    -\frac{v_\chi}{v_\varphi} (h^{f T})^{-1} Y^{f T} & 
    1 & 
    0 & 
    0 \\
    -\frac{v_\eta}{v_\varphi} (h^{\mathcal{F} T})^{-1} y^{\mathcal{F} T} & 
    0 & 
    1 & 
    0 \\
    0 & 0 & 0 & 1
\end{pmatrix}, 
\end{aligned}
\end{equation}
}

As a result, three of the dark neutral leptons acquire see-saw-suppressed masses originating from the $m_f$ matrix and are significantly lighter than the others. The lightest of these fermions serves as the DM candidate.

\subsection{Quark Spectrum}

We now proceed to apply a similar analysis to the quark sector, detailing the corresponding Yukawa interactions. The general Yukawa Lagrangian for the quarks is written as:
\be \label{eq:QuarkYuk}
\begin{split}
-\mathcal{L}_{q} &= Y^d_{aj}\,\overline{Q_{aL}}\, \eta^*\, d_{jR} 
+ Y^D_{ab}\,\overline{Q_{aL}}\, \chi^*\, D_{bR} 
\\
&+ Y^u_{3j}\,\overline{Q_{3L}}\, \eta\, u_{jR} 
+ Y^U_{33}\,\overline{Q_{3L}}\, \chi\, U_{3R}  
\\
&+ y^{\mathcal{K}}_{ab}\,\overline{Q_{aL}}\, \eta\, \mathcal{K}_{bR} + y^{\mathcal{K}}_{33}\,\overline{Q_{3L}}\, \eta^*\, \mathcal{K}_{3R} 
\\
&+ y^u_{aj}\,\overline{\mathcal{K}_{aL}}\, \chi\, u_{jR} 
+ y^d_{3i}\,\overline{\mathcal{K}_{3L}}\, \chi^*\, d_{iR}  \\
& + h^{\mathcal{K}}_{ab}\, \varphi\, \overline{\mathcal{K}_{aL}}\, \mathcal{K}_{bR} 
+ h^{\mathcal{K}}_{33}\, \varphi\, \overline{\mathcal{K}_{3L}}\, \mathcal{K}_{3R} 
+ \textrm{h.c.}\,, 
\end{split}
\ee

where $a,b = 1,2$ and $i,j = 1,2,3$.

\subsubsection{Up-type Quarks}

Starting with the $P_M$-even up-type quarks, we obtain the following mass matrix when written in the basis $\textbf{u} = (u_{i}, \mathcal{U}^\prime_a)$:
\bea \label{eq:upmassM}
\mathcal{M}_{u} = 
\frac{1}{\sqrt{2}}\begin{pmatrix}
   0_{[2\times 3]} & y^{\mathcal{K}}_{[2\times 2]} v_\eta \\
   Y^{u}_{[1\times 3]} v_\eta & 0_{[1\times 2]} \\
   y^{u}_{[2\times 3]} v_\chi & h^{\mathcal{K}}_{[2\times 2]} v_\varphi
\end{pmatrix}\,.
\eea
For the sake of clarity, since the quark Yukawa couplings have different dimensions, we explicitly write their dimensions in the mass matrices throughout this section.
 
The hierarchy between the blocks allows us to employ the see-saw approximation used in the previous section. We do not provide all the details here, as the procedure is quite similar to the one applied in the lepton sector. After performing the block-diagonalization, we obtain the following two matrices:

\bea
\begin{aligned}
&(m^{2}_u)_{\mathrm{light}} \\
&\simeq
\frac{v_\eta^2}{2} 
\begin{pmatrix}
    \kappa_\chi^2 A \left[1 - Y^{u\,\dagger}(Y^{u} Y^{u\,\dagger})^{-1} Y^{u}\right] A^\dagger & 0_{[2\times 1]}\\
    0_{[1\times 2]} & Y^u Y^{u\,\dagger}
\end{pmatrix}\,, 
\end{aligned}
\eea

and  
\bea
(M^{2}_u)_\text{heavy} &\simeq& \frac{v_\varphi^2}{2} h^{\mathcal{K}} h^{\mathcal{K}\,\dagger} \,, 
\eea
where $A_{[2 \times 3]} = y^\mathcal{K}\, h^{\mathcal{K}\,-1}\, y^u$ and $\kappa_\chi = v_\chi / v_\varphi$.  

Consequently, two of the three light up-type quarks acquire see-saw-suppressed masses, proportional to $v_\eta \kappa_\chi$. Meanwhile, the remaining quark has a mass proportional to the electroweak scale, $v_\eta$, and is identified as the top quark.

We can determine a constraint on the $Y^u$ matrix related to the top quark. According to \cite{ParticleDataGroup:2024cfk}, the top quark mass is $172.57 \pm 0.29$ GeV. Thus, we have:
\begin{equation}
\sqrt{\frac{v_\eta^2}{2} Y^u (Y^u)^\dagger} =  \frac{v_\eta}{\sqrt{2}} \sqrt{Y_{1}^{u 2} + Y_{2}^{u 2} + Y_{3}^{u 2}} \sim 172.57,
\end{equation}
from which we obtain:
\begin{equation}
Y_t \equiv \sqrt{Y_{1}^{u 2} + Y_{2}^{u 2} + Y_{3}^{u 2}} \sim 1.
\end{equation}
The Yukawa coupling is constrained to be less than $1.67$ at $95\%$ confidence level (C.L.) \cite{CMS:2019unu}.

Regarding the $P_M$-odd up-quarks, we choose $\mathcal{\textbf{U}}=(U_{3}, \mathcal{U}_{3}, \mathcal{U}_{a})$ as the basis. In this framework, the mass matrix for the dark up-type quarks is given by:
\bea \label{eq:DarkupmassM}
\mathcal{M}_{u}^{\text{dark}} = \frac{1}{\sqrt{2}}
\begin{pmatrix}
   Y^{U}_{33}\,v_\chi & -y^{\mathcal{K}}_{33}\,v_\eta  & 0_{[1\times2]} \\
   0 & h^{\mathcal{K}}_{33}\,v_\varphi & 0_{[1\times2]}\\
   0_{[2\times1]} & 0_{[2\times1]} & h^{\mathcal{K}}_{[2 \times 2]}\, v_\varphi
\end{pmatrix}\,.
\eea 
It is clear that $\mathcal{U}_{a}$ does not mix with the other quarks, and their mass matrix corresponds to the bottom-right block of $\mathcal{M}_{u}^{\text{dark}}$. For the remaining two states, there is a small mixing of order $\mathcal{O}(v_\eta/v_\varphi)$, and their masses are approximately given by the upper diagonal entries of $\mathcal{M}_{u}^{\text{dark}}$.

\subsubsection{Down-type Quarks}

For the $P_M$-even down-type quarks, we choose $\textbf{d} = (d_{i}, \mathcal{D}^\prime_{3})$ as the basis and derive the mass matrix as follows:
\bea \label{eq:downmassM}
\mathcal{M}_{d} 
=\frac{1}{\sqrt{2}}
\begin{pmatrix}
   Y^{d}_{[2\times 3]}\,v_\eta & 0_{[2\times 1]}\\
   0_{[1\times 3]} & -y^{\mathcal{K}}_{33}\,v_\eta\\
   y^{d}_{[1\times 3]}\,v_\chi & h^{\mathcal{K}}_{33}\,v_\varphi
\end{pmatrix}\,.
\eea

Following the see-saw procedure, the diagonalization of $\mathcal{M}_d \mathcal{M}_d^\dagger$ proceeds analogously to the lepton case. The masses are approximately given by:  
\begin{equation}
\begin{split}\label{eq:downdiagmasses}
&(m^{2}_d)_\mathrm{light} \\
&\simeq
\frac{v_\eta^2}{2} 
\begin{pmatrix}
    Y^d Y^{d\,\dagger} & 0_{[2\times 1]}\\
    0_{[1\times 2]} & \kappa_\chi^2\,B \left[1 - Y^{d\,\dagger}(Y^d Y^{d\,\dagger})^{-1}Y^d\right] B^\dagger
\end{pmatrix}\, , \\
&(M^{2}_d)_\mathrm{heavy} \simeq \frac{v_\varphi^2}{2} |h^{\mathcal{K}}_{33}|^2 \,,  
\end{split}
\end{equation}  
where $B_{[1 \times 3]} = \frac{y^\mathcal{K}_{33}}{h^\mathcal{K}_{33}}\,y^d$. As a result, among the down-type quarks, only one acquires a see-saw-suppressed mass.

Finally, we considered the $P_M$-odd down-quarks. To analyze this, we fix the basis for the dark down-type quarks as $\textbf{D} = (D_{a}, \mathcal{D}_{a}, \mathcal{D}_{3})$. The resulting mass matrix is given by:
\bea \label{eq:DarkdownmassM}
\mathcal{M}_d^{\text{dark}} =
\frac{1}{\sqrt{2}}
\begin{pmatrix}
   Y^{D}_{[2\times 2]}\,v_\chi & y^{\mathcal{K}}_{[2\times 2]}\,v_\eta & 0_{[2\times 1]} \\
   0_{[2\times 2]} & h^{\mathcal{K}}_{[2\times 2]}\,v_\varphi & 0_{[2\times 1]}\\
   0_{[1\times 2]} & 0_{[1\times 2]} & h^{\mathcal{K}}_{33}\,v_\varphi
\end{pmatrix}\,.
\eea 
The diagonal entries provide the approximate masses. While $\mathcal{D}_{3}$ remains unmixed, the fields $D_{a}$ and $\mathcal{D}_{a}$ exhibit a small mixing suppressed by $\kappa_\eta=v_\eta / v_\varphi$.

In summary, this section demonstrated that all fermions acquire mass through a consistent application of the see-saw mechanism, highlighting its universality across the entire fermion sector.

\subsubsection{LHC phenomenology}
The vector-like quarks $\mathcal{U}_a, \mathcal{U}_3, \mathcal{U}'_3, \mathcal{D}_a, \mathcal{D}_3$ and $\mathcal{D}'_3$ have masses proportional to $v_\varphi$ and are therefore far too heavy to be produced at the LHC. The 331-like exotic quarks $U_3$ and $D_a$, however, are relatively light, with masses proportional to $v_\chi$. They are therefore within the reach of LHC searches. %The LHC phenomenology of quarks is essentially that of 331-models, as the vector-like quarks are decoupled due to their large mass. 
The main LHC production channel for the exotic quarks is the gluon-gluon fusion, the exotic quarks being produced in pairs, $gg\to J\bar{J}$, or in association with an exotic gauge boson $gg\to V^{\pm} J, V^0 J$, where $J=U_3, D_a$. The decay modes of exotic quarks differ from 331 models \cite{Cao:2016uur}, due to the matter parity $P_M$. The exotic quarks $U_3$ and $D_a$ are odd under the matter parity and therefore ultimately decay into dark matter and SM particles. The main decay channels are: $U\to b V^{+\ast}\to b f_d X$ and $D_a\to (u/c) V^{-\ast}\to (u/c)f_d X$, where $X$ are leptonic SM final states. The LHC searches of exotic up- and down-type quarks impose the following tentative bound on their masses: $m_U>1.6$ TeV and $m_{D_a}>1.1$ TeV \cite{CMS:2022fck, ATLAS:2024gyc}.

\section{Bounds from Charged Leptons} \label{sec:bounds}

The flavour-violating Yukawa couplings are obtained from Eq. (\ref{eq:LepYuk}).  
After the block-diagonalization, keeping the leading order, the 
charged lepton Yukawa couplings can be divided into pure SM lepton couplings

\be
\begin{split}\label{eq:SMLepYuk}
\mathcal{L}_{\textrm{SM}} &=\frac{\sqrt{2}}{v_\eta}\bar{e}_L\, M_e\,e_R \eta_1^{0\ast}
+\frac{\sqrt{2}}{v_\chi}\bar{e}_L\, M_e\, e_R \chi_3^{0\ast}
\\
&-\frac{\sqrt{2}}{v_\varphi}\bar{e}_L\, M_e\, e_R \varphi+\hc,
\end{split}
\ee

and couplings containing exotic charged leptons
\be
\begin{split}\label{eq:ExoticLepYuk}
\mathcal{L}_{\rm exotic}
&=-\bar{e}_L \,y^F\, E'_R \eta^{0\ast}_1
+\bar{e}_L \,y^F\, E_R \eta^{0\ast}_3
\\
&+\bar{E}_L \,y^e\, e_R \chi^{0\ast}_1+\bar{E}'_L \,y^e\, e_R \chi^{0\ast}_3
\\
&+\bar{E}_L \,h^F\, E_R \varphi+\bar{E}'_L \,h^F\, E'_R \varphi +\hc
\end{split}
\ee
The charged lepton mass matrices in Eqs. (\ref{E-lepton mass matrix}) and (\ref{e and Ep mass matrices}) are diagonalized in the form $U^{l\dagger}_{L}M_l U^l_R=M_l^{\rm{diag}}$ and physical coupling constants in Eq. (\ref{eq:ExoticLepYuk}) become
\be
\begin{split}
&y^{\mathcal{F}}\to\widetilde{y}^{\mathcal{F}}=U^{e\dagger}_L y^{\mathcal{F}}U^E_R,\quad
y^{e}\to\widetilde{y}^{e}=U^{E\dagger}_L y^{e}U^e_R,\\
&h^{\mathcal{F}}\to\widetilde{h}^{\mathcal{F}}=U^{E\dagger}_L h^{\mathcal{F}}U^E_R.
\end{split}
\ee
One can see that the Yukawa couplings of SM charged-leptons are proportional to their mass matrix and are therefore diagonalized simultaneously with the SM lepton mass matrix.
Therefore the neutral scalars do not have flavour-changing neutral currents at the tree level. This also avoids the LHC bounds on the lepton flavour-violating couplings of Higgs. The SM lepton couplings are all suppressed by heavy VEVs, except the first one in Eq.~\eqref{eq:SMLepYuk}. The muon and tau Yukawa couplings have been measured at LHC to be close to the SM values. Therefore the real part of $\eta_1^0$ needs to be identified with $125$ GeV Higgs. The lepton masses are suppressed by $v_\chi/v_\varphi$. This imposes $v_\chi/v_\varphi\gtrsim 10^{-2}$, in order to keep Yukawa coupling smaller than 1.   

\begin{figure}[th!]
    \centering
    \scalebox{0.7}{\dominantScalar  \hspace{0.7cm}
    \dominantGauge}
    \caption{The dominant diagrams contributing to $\mu\to e\gamma$.}
    \label{dominant LFV diagrams}
\vspace*{-2mm}
\end{figure}

\begin{figure}[ht]
    \centering
    \scalebox{0.7}{
    \ComboOne  
    \hspace{0.2cm}
    \ComboTwo 
    }
    \\
    \scalebox{0.7}{
    \ComboThree  
    \hspace{0.2cm} 
    \ComboFour
    }
    \\
    \scalebox{0.7}{
    \ComboFive  
    \hspace{0.2cm}
    \ComboSix
    }
    \caption{Sub-leading diagrams contributing to $\mu\to e\gamma$. }
    \label{subdominant LFV diagrams}
\vspace*{-2mm}
\end{figure}

Even though scalars do not have flavour-changing couplings to a pair of SM charged leptons,  the flavour-violating  effects appear at the loop level due to the mixing between SM leptons and the charged exotic leptons. The most constraining charged lepton flavour-violating (CLFV) process is $\mu\to e\gamma$, where all neutral scalars and  massive gauge bosons mediate it at loop level. 
 The most constraining  CLFV diagrams are presented in Fig.~\ref{dominant LFV diagrams} and sub-leading diagrams in Fig.~\ref{subdominant LFV diagrams}. Note that the photon vertex is always flavour conserving.
The couplings between scalars, SM charged leptons and $E'_i$ are not suppressed by the mixing between SM leptons and exotic leptons, as can be seen in Eq. (\ref{eq:ExoticLepYuk}). The scalar mediated diagrams are, however, heavily suppressed by large masses of $E'_i$ and $E_i$, as well as the relation 
\begin{equation}
\widetilde{y}^{\mathcal{F}}(M_{E'}^{\rm{diag}})^{-1}\widetilde{y}^e=\frac{2}{v_\eta v_\chi}M_e^{\rm{diag}},
\end{equation}
which originates from the see-saw mechanism. 
The leading contribution to the scalar diagrams  is cancelled due to this relation, providing further suppression. The most relevant scalar diagram is mediated by $H_3$, corresponding to the branching ratio
\begin{equation}
\begin{aligned}
\textrm{BR}(\mu\to e\gamma) &= \frac{e^2 \alpha^2_{13}\alpha^2_{23}m_\mu^3 m_{H_3}^4}{64(4\pi)^5 \Gamma_{\mu}^{\textrm{tot}}} \\
&\quad \times \left( \left|\sum_{i=1}^3\frac{\widetilde{y}^e_{ie}\widetilde{y}^F_{\mu i}}{m_{E'_i}^3}\right|^2 + \left|\sum_{i=1}^3\frac{\widetilde{y}^e_{i\mu}\widetilde{y}^F_{e i}}{m_{E'_i}^3}\right|^2 \right),
\end{aligned}
\end{equation}
where $\Gamma_\mu^{\rm{tot}}$ is the total muon decay width. 
We can obtain a conservative bound by setting the $m_{H_3},m_{E'_i}\sim v_\varphi$ and the flavour-violating couplings to unity:
\begin{equation}
\left(\frac{\rm{GeV}}{v_\varphi}\right)\alpha_{13}\alpha_{23}<1.55\times 10^{-10},
\end{equation}
which is satisfied for $v_\varphi>10^6 \rm{GeV}$ and $\alpha_{13},\alpha_{23}<10^{-2}$. So even with large, $\mathcal{O}(1)$, flavour-violating couplings, bound from this diagram can be avoided.  

The diagrams mediated by the exotic gauge bosons are suppressed by their heavy masses. The couplings involved are also tiny, $\sim v_\eta/v_\varphi$, due to them being a result of mixing between SM leptons and exotic ones. These diagrams provide an insignificant contribution to $\mu\to e\gamma$.
The diagrams mediated by $W$ boson are suppressed due to tiny $\nu_i$ masses or small mixing between heavy and light neutrinos. The only relevant diagram mediated by gauge bosons is the $Z$ boson mediated one in Fig.~\ref{dominant LFV diagrams}. It is suppressed only by one flavour-violating coupling and is not mass suppressed due to $Z$ boson being relatively light. 
It provides contribution comparable to $H_3$ mediated diagram. The $\mu\to e\gamma$ branching ratio corresponding to $Z$ boson mediated diagram is,
\be
\begin{split}
&\textrm{BR}(\mu\to e\gamma)\\
&=\frac{e^2 g^4 m_\mu^5 v_\chi^4 \sin^2\theta_W\tan^2\theta_W}{144(4\pi)^5 \Gamma_\mu^{\textrm{tot}}m_z^4}\left|\frac{3g_3^2}{g_x^2}\Lambda^{\ast}_{\mu e}+3\Lambda_{e\mu}\right|^2,
\end{split}
\ee
where
\be
\Lambda=\widetilde{y}^{e\dagger}((m_{E'}^{\rm{diag}})^{-1})^2 \widetilde{y}^e.
\ee
We can obtain a conservative bound by setting $\Lambda\sim 1/v_\varphi^2$
\be
\frac{v_\chi}{v_\varphi}<7.7\times 10^{-3}.
\ee

Besides \(\mu\!\to e\gamma\), our setup induces \(\mu\!\to\!3e\) and coherent \(\mu\!-\!e\) conversion once the heavy states are integrated out. The electromagnetic dipole controlling \(\mu\!\to e\gamma\) receives contributions from both the \(H_3\)–\(E_i'\) loop (Eqs.~(80)–(81)) and the \(Z\) penguin (Eqs.~(82)–(83)), while the seesaw relation (Eq.~(79)) cancels the leading scalar pieces. In the dipole-dominance limit one finds the standard relation
\be
\begin{split}
\mathrm{BR}(\mu\!\to\!3e) & \;\simeq\;\frac{\alpha}{3\pi}\!\left[\ln\!\frac{m_\mu^2}{m_e^2}-\frac{11}{4}\right]\!
\mathrm{BR}(\mu\!\to e\gamma)\\
&\;\approx\; 6.13\times10^{-3}\,\mathrm{BR}(\mu\!\to e\gamma),
\end{split}
\ee
in agreement with Eq.~(117) of Ref.~\cite{Kuno:1999jp}. This implies that current \(\mu\!\to\!3e\) bounds are weaker than those from \(\mu\!\to e\gamma\). Coherent \(\mu\!-\!e\) conversion probes the same dipole coefficients together with vector/scalar operators generated by \(Z,Z',Z''\) penguins and neutral-scalar exchange; in the parameter region analyzed here (heavy \(Z',Z''\) and \(H_{1,2,4}\), seesaw-suppressed \(e\text{--}E'\) mixing) the present limits are not more restrictive than \(\mu\!\to e\gamma\). We stress that the dark-matter benchmarks used in Sec.~VIII adopt diagonal Yukawa textures, further suppressing new LFV sources and rendering our benchmark points conservative.

In conclusion the CLFV constraints are easily avoided, even with large flavour-violating couplings. As mentioned in Section \ref{sssec:cleptons}, in our numerics for the dark matter phenomenology we made the simplifying assumptions of making the Yukawa couplings diagonal. In that case the CLFV effects are absent.

\section{Dark Matter} \label{sec:dm}
As previously mentioned, after the symmetry breaking of the $3$-$3$-$1$-$1$ gauge symmetry, the $P_M$ symmetry naturally emerges within the model. This discrete symmetry plays a crucial role in stabilizing the lightest $P_M$-odd particle, which, being electrically neutral, serves as a compelling DM candidate. This particle could account for the DM relic abundance measured by the Planck satellite \cite{Planck_2018}, given by:
\begin{equation*}
\Omega_{\textrm{DM}} h^{2} = 0.1200 \pm 0.0012. \label{relic_density}
\end{equation*}
The model predicts the existence of neutral $P_M$-odd particles across all sectors: scalar, fermion, and gauge. Specifically, these include the scalar $S_d$ in Eq.~\eqref{Sd_scalar}, the gauge boson $V^{0}$ in Eq.~\eqref{charged_gauge_bosons}, and the leptons, which are the eigenstates of the mass matrices defined in Eq.~\eqref{dark_leptons}. However, as $P_M$ is a $Z_2$ discrete symmetry, the model does not support a multi-component DM sector.

The $V^{0}$ gauge boson is not a suitable DM candidate because its mass, given by $m_{V^{0(\dagger)}}^2 = m_{W^\pm}^2 + m_{V^\pm}^2$, allows it to decay into $W^\pm$ and $V^\mp$. Since $V^\pm$ is also a $P_M$-odd particle, this decay renders $V^{0}$ unstable, disqualifying it as a viable DM candidate. As for the scalar candidate, $S_d$, while it could theoretically provide the correct relic density, it is completely ruled out by current experimental constraints on the spin-independent direct detection cross-section, which exclude all regions of its parameter space. Consequently, we will not consider it further in this work.

On the other hand, the dark fermion sector offers a promising candidate for DM. Specifically, the lightest eigenstate of the mass matrix $m_f$ in Eq.~\eqref{dark_leptons}, which we denote as $f_{\textrm{d}}$, emerges as a viable DM candidate. The remaining dark leptons, i.e. the eigenstates of $m_{f_1^{'}}$, $m_{f_2^{'}}$, and $m_{f_3^{'}}$, are extremely heavy, with masses proportional to $v_\varphi \sim 10^{6}$ GeV. Consequently, they do not contribute to the DM phenomenology and are integrated out in the effective Lagrangian.

In fact, not only are the heavy dark leptons integrated out, but all particles with masses on the order of $v_\sigma$ and $v_\varphi$ are as well, as they do not play a significant role in the DM analysis. This is because the focus is on DM candidates with masses in the range of 10 GeV to 1.5 TeV, a range that aligns with the sensitivity of current and upcoming accelerator experiments.

Before detailing the parameter scan, an important methodological clarification is in order. Although one could in principle integrate out heavy states sequentially and run the RGEs between the four breaking scales $v_\eta$, $v_\chi$, $v_\sigma$ and $v_\varphi$, we deliberately adopt a single one-loop potential evaluated at a single renormalisation scale. The justification is twofold:
\begin{enumerate}

    \item[(i)]  The fields that gain masses at the highest scales, $v_\sigma$ (associated with $U(1)_N$ breaking) and $v_\varphi$ (scale-invariance breaking), lie in the $\mathcal{O}(10^{2}\text{–}10^{3})$ TeV range and therefore decouple completely from TeV phenomenology and from the dark-matter freeze-out epoch. We integrate them out at tree level and keep their leading loop effects through the Gildener-- Weinberg formalism.
\item[(ii)] The residual hierarchy $v_\chi/v_\eta\lesssim10^{2}$ generates at most
\[
\frac{1}{16\pi^{2}}\,
\ln\!\bigl(v_\chi^{2}/v_\eta^{2}\bigr)\;\lesssim\;0.06 ,
\]
so even $\mathcal{O}(1)$ couplings shift the light-sector quartics by only a few per cent. This shift in the light-sector quartics, by only a few per cent ($\lesssim 0.06$), is below our scan granularity. It is also modest when compared to the broader theoretical and phenomenological uncertainties inherent in dark matter calculations, which can often significantly exceed the ${\sim}1\%$ numerical precision typically targeted by tools like {\tt MicrOMEGAs} for standard computations~\citep{ALGUERO2024109133}.
\end{enumerate}

Hence the single-step approach provides accuracy sufficient for the dark-matter analysis, while a full multi-threshold RG study—though feasible—would add substantial technical overhead without altering our conclusions and is deferred to future work.

To analyze the phenomenology of DM, we begin by fixing certain parameters of the model. We first focus on the lepton Lagrangian, where the relevant parameters are \( Y^\nu \), \( Y^f \), \( h^{\nu} \), \( h^{f} \), \( y^\mathcal{F} \), \( y^e \), and \( h^\mathcal{F} \). All these quantities are \(3 \times 3\) matrices.

Furthermore, throughout the dark matter analysis, we maintain the CP-conserving setup: all Yukawa couplings and scalar potential parameters are real. This ensures that relic density predictions are free from additional CP-phase dependencies and automatically satisfy current EDM bounds, while remaining robust against the anticipated two-orders-of-magnitude sensitivity improvements from forthcoming molecular-lattice experiments such as BaOH~\cite{BaOH2024}.

The matrices \( h^\mathcal{F} \), \( h^{\nu} \), and \( h^{f} \) are required to be of order one to preserve the validity of the see-saw mechanism in both the charged and neutral lepton sectors. For simplicity, we assume $h^\nu$ and $h^f$ matrices to be equal to the identity matrix and $h^{\mathcal{F}}$ to be $0.5$ times the identity matrix. Similarly, the \( y^\mathcal{F} \) matrix is also taken as the identity matrix. The \( y^e \) matrix, in contrast, is diagonal, with its elements specified by Eq.~\eqref{ye_matrix}, ensuring compatibility with the SM charged leptons. The \( Y^\nu \) matrix is derived using Eq.~\eqref{Y_nu_matrix} to reproduce the active neutrino masses and mixings, consistent with neutrino oscillation data \cite{esteban2024nufit60updatedglobalanalysis}. 

Finally, the Yukawa coupling matrix \( Y^f \), assumed to be diagonal for simplicity, is constrained by \( Y^f < \sqrt{4 \pi} \, \mathbb{1} \) to maintain the perturbative regime. The diagonal elements of this matrix are directly linked to the masses of the dark leptons, \( m_f \), as specified in Eq.~\eqref{dark_leptons}.

In a similar manner, we fixed the parameters in the quark sector, which include the matrices \( Y^d_{aj} \), \( Y^D_{ab} \), \( Y^u_{3j} \), \( Y^U_{33} \), \( y^{\mathcal{K}}_{ab} \), \( y^{\mathcal{K}}_{33} \), \( y^u_{aj} \), \( y^d_{3i} \), \( h^\mathcal{K}_{ab} \), and \( h^\mathcal{K}_{33} \), where \( a, b = 1, 2 \) and \( i, j = 1, 2, 3 \). Among these, \( Y^d \) and \( Y^u \) are determined to reproduce the SM quark masses. For simplicity, the remaining matrices are chosen to be equal to the identity. This assumption is consistent with the see-saw mechanism, which is also present in the quark sector. It is important to emphasize that exotic quarks, with masses on the order of \( v_\chi \), do not significantly influence the DM analysis. 

At this point, we have fixed the parameter space in the scalar sector. Equation~\eqref{lambda phi} enables all quartic couplings to be expressed in terms of the scalar masses and rotation angles, as defined in Eq.~\eqref{O matrix}. Additionally, three of these angles have been reformulated in terms of the VEVs, as shown in Eq.~\eqref{alphas_123}. Based on the previous discussion, the free parameters used in our DM analysis are: $\alpha_{12}$, $\alpha_{13}$, $\alpha_{23}$, $m_{H_2}$, $m_{H_3}$, $v_\chi$, $v_\sigma$, and $v_\varphi$. Specifically, the ranges and relationships are defined as follows. For the mixing angles, the allowed intervals are:
\begin{equation} 
\begin{split}
10^{-4} &\leq \abs{\alpha_{12}} \leq 10^{-1}, \\ 
10^{-4} &\leq \abs{\alpha_{13}} \leq 10^{-3},
\\
10^{-4} &\leq \abs{\alpha_{23}} \leq 10^{-2}.
\label{eq:angles}
\end{split}
\end{equation}
The angles, expressed in radians, are, in principle, selected to ensure that all quartic couplings comply with the perturbative unitarity constraints discussed in Sec.~\ref{sec:pert}. However, the maximum value of $\alpha_{12} = 0.32$ is already excluded by direct detection bounds. Therefore, we adopt a more conservative upper limit of $10^{-1}$.

The vacuum expectation values are constrained by:
\begin{equation}
\begin{split}
    10 \, \mathrm{TeV} &\leq v_\chi \leq 20 \, \mathrm{TeV}, \\
    10^2 \, \mathrm{TeV} &\leq v_\sigma \leq 1.5 \cdot 10^3 \, \mathrm{TeV}, \\
    10^3 \, \mathrm{TeV} &\leq v_\varphi \leq 2.5 \cdot 10^3\, \mathrm{TeV}.
\label{eq:vevs}
\end{split}
\end{equation}
Here it is important to note that although the minimum $v_\chi$ value allowed by eq. \eqref{eq:lower_bound_vchi} is approximately 3.6 TeV, we conservatively adopt a minimum value of 10 TeV, guided by the conservative $Z^{\prime}$ mass lower bounds established in recent studies \cite{PhysRevD.106.055027}. While those studies focused on specific 331 models distinct from ours, their results (exceeding 4 TeV from current LHC data, and projected above 5.8 TeV at HL-LHC, 9.9 TeV at HE-LHC, and 27 TeV at FCC, considering exotic/invisible decays) provide strong support for our conservative choice.

Finally, the masses of the scalar particles are given by:
\begin{equation} 
    100\, \mathrm{GeV} \leq m_{H_2} \leq  2 \, \mathrm{TeV}, \quad
10\, \mathrm{TeV} \leq m_{H_3} \leq 100 \, \mathrm{TeV}.
\label{eq:masses}
\end{equation}
At this point, an important clarification regarding our choice of scanning variables is in order. Our analysis rigorously preserves classical scale invariance.  
The scalar quartic couplings $\lambda_i$, the fundamental dimensionless inputs of the theory, are \emph{reparametrised} via eq.~\eqref{lambda phi} in terms of  (i) the dimensionless mixing angles $\alpha_{ij}$,  
(ii) the physical CP-even scalar masses $m_{H_i}$, and  
(iii) the vacuum-expectation values (vevs) $v_i$.  
This change of variables is \emph{bijective} within the stable, perturbative region of the potential, hence it is mathematically equivalent to scanning directly over the $\lambda_i$.

Crucially, $m_{H_i}$ and $v_i$ are not free inputs: at each sampled point they satisfy the minimisation and flat-direction conditions required by the Coleman–Weinberg mechanism.  The reparametrisation therefore serves only to make the scan outputs more transparent, because $\alpha_{ij}$, $m_{H_i}$ and $v_i$ map directly onto observables such as Higgs masses and electroweak vevs.  
All other dimensionless couplings—Yukawas $(h^{\mathcal F},\,Y^{f},\ldots)$ and gauge couplings $(g,\,g_X,\ldots)$—are scanned explicitly as fundamental inputs.  
Hence the procedure remains fully consistent with scale invariance while streamlining the connection to experimental constraints.

With all parameters established, we now proceed to calculate the DM relic density, $\Omega_{\textrm{DM}}h^{2}$, resulting from the freeze-out of $f_{\textrm{d}}$. To achieve this, we follow the standard procedures outlined in Refs.~\cite{gondolo1991,griest1991}. These procedures involve solving the Boltzmann equation for the DM yield $Y$, defined as $Y \equiv n / s$ (the ratio of number density to entropy density), which is expressed as:

\begin{equation} \label{Boltzmann_eq}
    \frac{dY}{dx} = -\left(\frac{45}{\pi}G\right)^{-1/2} 
    \frac{g_{*}^{1/2} m_{f_\textrm{d}}}{x^{2}} 
    \langle \sigma v \rangle_{\textrm{ann}} \left(Y^{2} - Y_{\textrm{eq}}^{2}\right).
\end{equation}
Here, $x = m_{f_\textrm{d}} / T$ represents the dimensionless temperature variable, $G$ denotes the gravitational constant, and $Y_{\textrm{eq}} = n_{\textrm{eq}} / s$ is the equilibrium yield. The equilibrium number density, \( n_{\textrm{eq}} \), and the effective number of relativistic degrees of freedom, \( g_{*} \), are defined following the formalism outlined in Ref.~\cite{gondolo1991}. 
Finally, $\langle \sigma v \rangle_{\textrm{ann}}$ corresponds to the thermally averaged annihilation cross section, a key quantity in the freeze-out calculation. Once the Boltzmann equation is solved and the DM yield $Y$ is determined at the present time ($Y_0$), the relic density is computed as:
\begin{equation}
    \Omega_{\textrm{DM}} h^{2} = 2.82 \times 10^{8} \times Y_{0} \times \frac{m_{f_\textrm{d}}}{\textrm{GeV}}.
\end{equation}

\begin{figure}[!ht]
    \centering
    \scalebox{0.7}{
    \DMZWW 
    \hspace{0.2cm}
    \DMVee 
    }
    \\
    \scalebox{0.7}{
    \DMHee  
    \hspace{0.2cm} 
   \DMHWW
    }
    \caption{Dominant annihilation channels for the DM fermion $f_\mathrm{d}$ into Standard Model particles include: $f_{\rm d}\bar{f}_{\rm d} \to Z^{*} \to W^+W^-$; $f_{d}\overline{f}_{d}\rightarrow e^{+}e^{-}$ (t-channel $V^{\pm}$ exchange); and $f_{\rm d}\bar{f}_{\rm d} \to H_{1,2} \to e^+e^-, b\bar{b}, W^+W^-$.}
    \label{ffee diagrams}
\vspace*{-2mm}
\end{figure}

Due to the complexity of the model and the extensive parameter space fixed by
eqs.~\eqref{eq:angles}--\eqref{eq:masses}, solving the Boltzmann
equation~\eqref{Boltzmann_eq} is computationally demanding. The main
difficulty lies in evaluating the thermally-averaged annihilation rate
\(\langle\sigma v\rangle_{\text{ann}}\), which, in our setup, is governed by
two leading channels illustrated in Fig.~\ref{ffee diagrams}:
(i) \emph{Vector-portal processes}:  
\(f_{\mathrm{d}}\bar{f}_{\mathrm{d}} \to Z^{*} \to W^{+}W^{-}\) ($s$-channel)  
and  
$f_{d}\overline{f}_{d}\rightarrow e^{+}e^{-}$ ($t$-channel $V^{\pm}$ exchange);  
(ii) a \emph{scalar-portal} process, \(f_{\mathrm{d}}\bar{f}_{\mathrm{d}} \to H_{1,2} \to e^{+}e^{-}, b\bar{b}, W^{+}W^{-}\). These channels dominate the relic abundance by determining the value of $\langle\sigma v\rangle_{\text{ann}}$ that enters the Boltzmann analysis. To address this, we employ the packages {\tt Feynrules}~\citep{FeynRules_2014,cao2016colliderphenomenology331model}, {\tt Calchep}~\citep{BELYAEV20131729}, and {\tt MicrOMEGAs}~\citep{ALGUERO2024109133}, which are widely used tools in phenomenological studies of dark matter. Specifically, we used the {\tt MicrOMEGAs} package to scan over the DM mass in the range \( 10~\textrm{GeV} \leq m_{f_\mathrm{d}} \leq 1500~\textrm{GeV} \)  and varied the remaining parameters within the ranges defined in eqs.~\eqref{eq:angles}--\eqref{eq:masses}.

\begin{figure}[!ht] 
    \begin{center}
  \includegraphics[width=1\linewidth]{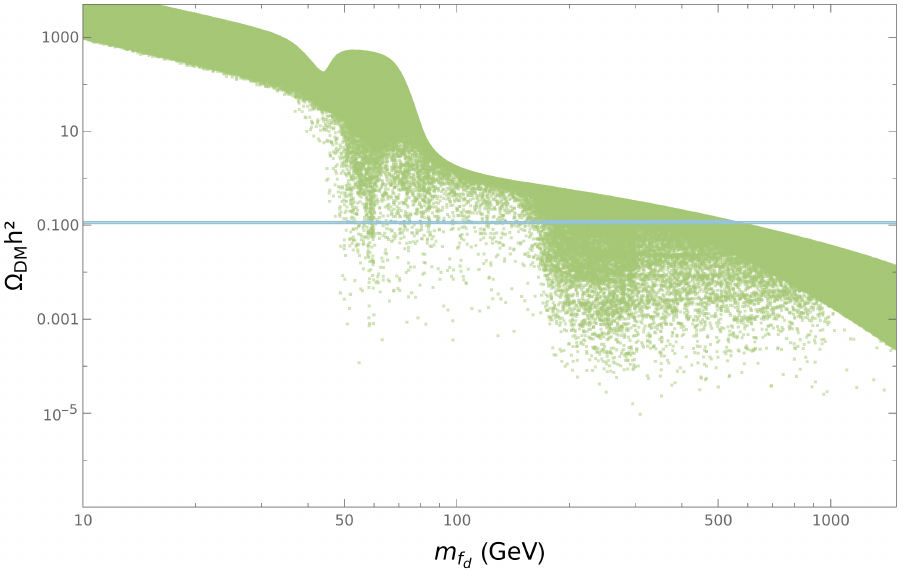}
    \end{center}
\begin{minipage}{1\linewidth}
        \caption{ Relic density $\Omega_{\textrm{DM}} h^{2}$ as a function of the fermionic dark matter mass $m_{f_\textrm{d}}$. The green points represent the results obtained from a logarithmic random scan over the model parameters. The blue horizontal band shows the measured relic density at the $3\sigma$ confidence level by the Planck collaboration~\cite{Planck_2018}. The figure illustrates the regions of parameter space satisfying the relic density constraint, particularly the viable mass window \(220~\textrm{GeV} \lesssim m_{f_\textrm{d}} \lesssim 555~\textrm{GeV}\) primarily driven by the $H_2$ resonance.} \label{fig:DDall3sig}
    \end{minipage}
\end{figure}

Our results are presented in Fig.~\ref{fig:DDall3sig}, where we show the relic density (green region) obtained from a logarithmic random scan over the parameters defined in eqs.~\eqref{eq:angles}-\eqref{eq:masses}. The horizontal band in this figure represents the measured relic density, $\Omega_{\textrm{DM}} h^{2}$, at the $3\sigma$ level, as reported by the Planck collaboration~\cite{Planck_2018}. It is evident that the fermionic dark matter candidate in this model satisfies the relic density constraint within the range \(220~\textrm{GeV} \lesssim m_{f_\textrm{d}} \lesssim 555~\textrm{GeV}\). For $m_{f_\textrm{d}} < 100$ GeV, the relic density exhibits typical resonant behaviour, showing the Z-boson resonance around 45 GeV and the Higgs resonance around 62.5 GeV. However, while these resonance regions can yield parameter points satisfying the relic density constraint, none of them manage to simultaneously satisfy the current direct detection limits. This is primarily because, to satisfy direct detection constraints, $f_\textrm{d}$ must be sufficiently decoupled from Standard Model particles. In the explored mass range for $f_\textrm{d}$, the $Z'$ gauge boson does not play a significant role because, as mentioned, its mass is set above 4 TeV. Specifically, within the range \(220~\textrm{GeV} \lesssim m_{f_\textrm{d}} \lesssim 555~\textrm{GeV}\), the annihilation cross-section $\langle \sigma v \rangle_{\text{ann}}$ is significantly enhanced around the $H_2$ resonance mass, becoming large enough to lower the relic density to the observed value and thus allowing for viable dark matter candidates. For $m_{f_\textrm{d}}$ masses larger than approximately $555$ GeV, the relic density falls below the observed value, leading to under-abundant dark matter. In this mass region, the dark matter phenomenology in the presented model is predominantly determined by the presence of the new scalar $H_2$.

\begin{figure}[ht!]
    \centering
\includegraphics[width=1\linewidth]{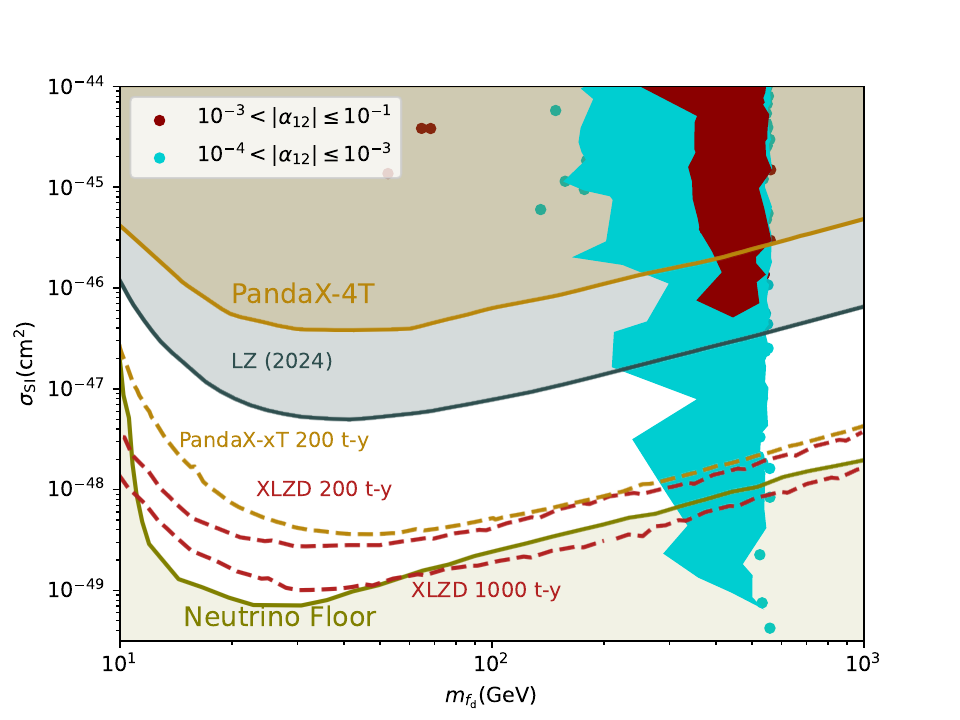}
    \begin{minipage}{1\linewidth}
    \caption{ Spin-independent direct detection results for the dark matter fermion $f_\textrm{d}$. The figure shows the elastic scattering cross section per nucleon, $\sigma_{\textrm{SI}}$, versus dark matter mass $m_{f_\textrm{d}}$. Parameter points satisfying the relic density constraint are shown, with different regions illustrating results corresponding to varying values of the $\alpha_{12}$ angle, as discussed in the text. Current experimental limits (PandaX-4T~\cite{PhysRevLett.127.261802}, LZ~\cite{LZ2024}), projected future sensitivities (XLZD~\cite{Aalbers_2022}, PandaX-xT~\cite{Abdukerim_2024}; dashed lines), and the neutrino floor~\cite{O_Hare_2021} are also displayed for comparison.}
    \label{fig:SI}
    \end{minipage}   
\end{figure}

Another important set of constraints on DM candidates comes from contemporary direct detection experiments~\citep{PhysRevLett.127.261802,LZ2024}, which aim to detect WIMP DM by measuring the kinetic energy transferred to a nucleus during its interaction with a DM particle. These experiments have established upper bounds on the WIMP-nucleus scattering cross section. In general, WIMP-nucleus interactions are classified as either spin-independent (SI) or spin-dependent (SD). Currently, the most stringent limit for spin-independent scattering is set at \(36~\textrm{GeV}/c^2\), rejecting cross sections above \(2.1 \times 10^{-48}~\textrm{cm}^2\) at the 90\% confidence level, as reported by the LUX-ZEPLIN (LZ) collaboration~\citep{LZ2024}. In this work, we analyse the DM candidate \(f_{\textrm{d}}\) and find that spin-independent (SI) interactions dominate. Using the {\tt MicrOMEGAs} package, we compute the SI elastic scattering cross section per nucleon, \(\sigma_{\textrm{SI}}\), for the parameter region that satisfies the relic density constraint. Fig.~\ref{fig:SI} presents the results for this region, showing different areas corresponding to varying values of the $\alpha_{12}$ angle, as this parameter plays a critical role in direct detection within the explored parameter space. It is clear that values of $\alpha_{12}$ greater than $10^{-3}$ are excluded by current direct detection limits, and a considerable region of the parameter space with $\alpha_{12}$ values in the range $10^{-4} < |\alpha_{12}| \le 10^{-3}$ is already eliminated by LZ~\citep{LZ2024}. Meanwhile, other significant region all into the neutrino floor, rendering them inaccessible to current direct detection experiments~\cite{O_Hare_2021}. Additionally, we present the sensitivity curves for the LZ \cite{PhysRevLett.131.041002} and PandaX-4T experiments \cite{PhysRevLett.127.261802}, along with the projected sensitivity curves for the near-future XLZD \cite{Aalbers_2022} and PandaX-xT \cite{Abdukerim_2024} experiments, shown as dashed lines. SD interactions are not considered in this analysis, as they are, in this model, several orders of magnitude weaker than the current experimental constraints on SD scattering.

\section{Conclusions}\label{sec:conc}
In this work, we propose a scale-invariant model based on the gauge group 
$\textrm{SU}(3)_{C} \otimes \textrm{SU}(3)_{L} \otimes \textrm{U}(1)_{X} \otimes \textrm{U}(1)_{N}$, 
with the $\textrm{U}(1)_{N}$ symmetry chosen such that a natural $P_M$ discrete symmetry emerges 
after the dynamical symmetry breaking. As part of the $3$-$3$-$1$ family, this construction has the potential 
to tackle several unresolved challenges of the SM, such as the origin of neutrino masses, the determination 
of the number of lepton and quark families, and the nature of dark matter.

One of the significant advances of the proposed framework is the reduction in the number of scalar fields 
in high representations of the gauge group, simplifying the scalar potential while preserving the ability 
to break the gauge symmetry to $\textrm{SU}(3)_{C} \otimes \textrm{U}(1)_{Q}$. This optimised scalar sector, 
combined with the imposition of scale invariance, enhances predictability compared to earlier 
$3$-$3$-$1$ constructions, which often suffered from excessive scalar complexity. In Sec.~\ref{scsect}, we analyse the 
scalar sector in detail, following the Gildener-Weinberg approximation, and demonstrate that, in addition to the 
SM-like Higgs, $H_1$, and the dilaton, $H_4$, the model contains two CP-even scalars, $H_{2,3}$, and one complex scalar, $S_{\textrm{d}}$. 

We found that the perturbative unitarity constraints, discussed in Sec.~\ref{sec:pert}, impose critical bounds on the quartic couplings and other parameters of the model, ensuring the theoretical consistency of the framework. By analysing two-to-two scattering amplitudes at high energy, we derived explicit conditions that all quartic couplings and eigenvalues of the scalar coupling matrix must satisfy. These constraints play a vital role in limiting the parameter space, directly influencing the scalar masses, mixing angles, and VEVs within the model. In particular, these conditions require the mixing angles $\alpha_{12}$, $\alpha_{13}$, and $\alpha_{23}$ to be small (see eq.~\eqref{eq:angles}).

Furthermore, in analogy to ref.~\cite{Dias:2022hbu}, the introduction 
of vector-like quarks and leptons, together with a natural choice of the different VEVs, partially addresses the hierarchy among the charged lepton 
and quark masses through a universal see-saw mechanism, facilitating the generation of masses 
for all SM particles as well as new exotic states at energy scales compatible with 
scale invariance. This framework surpasses the model proposed in ref.~\cite{Dias:2022hbu}, 
as it also provides an explanation for the hierarchy among the charged lepton masses, as shown in Sec.~\ref{scYukawa}. Charged lepton flavour-violating effects are present in the model due to mixing between SM charged leptons and the exotic leptons. For scalar mediated charged lepton flavour-violating processes the constraints are avoided due to heavy mass of exotic particles, as well as relations between the couplings due to see-saw mechanism. The gauge mediated charged lepton flavour-violating processes, on the other hand, are avoided due to small couplings.

The gauge sector in the scale-invariant 3-3-1-1 model, treated in Sec.~\ref{sec:gauge}, features a rich structure derived from the kinetic terms of the scalar bosons and their interactions with the gauge bosons. The charged (non-Hermitian) bosons arise from the non-diagonal generators of SU(3)$_L$ and acquire masses proportional to $v_\eta$ (which in this model is equal to $v_{\textrm{SM}}$) in the case of $W^{\pm}$, and proportional to the 3-3-1 energy scale, $v_\chi$. In the neutral boson sector, mixing among gauge fields produces four physical states: the massless photon and the $Z$, $Z'$, and $Z''$ bosons, whose masses are linked to the VEVs of the $\eta$, $\chi$, and $\sigma$ fields. Among these, the $Z''$ boson is particularly important as it stabilises the scalar potential, enabling dynamical symmetry breaking and ensuring the model's theoretical consistency at high energy scales. Furthermore, experimental bounds on the $\rho_0$ parameter, which measures the ratio of $W^\pm$ and $Z$ boson masses, impose a lower limit on the SU(3)$_L$ symmetry-breaking scale, $v_\chi \gtrsim 3.6 \, \text{TeV}$. This result underscores the model's compatibility with current data and its potential for future experimental validation. However, in our analysis, we adopted a more conservative value of 10 TeV for $v_\chi$, motivated by the experimental limits on the $Z'$ boson mass.

A key strength of the proposed model lies in its ability to naturally explain dark matter phenomenology. The model incorporates a dark matter candidate stabilised due to the $P_M$ symmetry, which emerges after symmetry breaking. Among the possible $P_M$-odd neutral particles, the fermion $f_\mathrm{d}$ is identified as a viable dark matter candidate. Our analysis shows that $f_\mathrm{d}$ satisfies the relic density constraint within the mass range $220 \, \textrm{GeV} \lesssim m_{f_\mathrm{d}} \lesssim 555 \, \textrm{GeV}$, as shown in Fig.~\ref{fig:DDall3sig}. This result primarily depends on the scalar mixing parameters $\alpha_{12}$ and the $v_\chi$. Spin-independent scattering cross-sections for $f_\mathrm{d}$ are consistent with experimental limits from LZ and PandaX-4T in certain regions of the parameter space (see Fig.~\ref{fig:SI}). Moreover, some parts of the parameter space lie below the neutrino floor, making them inaccessible to current detection technologies.

Other dark matter candidates, such as the vector boson $V^0$ and the scalar $S_d$, are excluded due to their instability or incompatibility with direct detection constraints. Furthermore, a substantial portion of the parameter space can be probed in upcoming experiments, such as XLZD and PandaX-xT, offering opportunities for further validation of the model's predictions.

\acknowledgments

We dedicate this work to the memory of Alex G. Dias, an esteemed colleague and dear friend, whose contributions to science and kindness deeply impacted those who worked alongside him. His legacy will continue to inspire us and future generations. We would like to thank Y. Villamizar, Y. Porto, J. P. Carvalho-Correa, and A. C. D. Viglioni  for discussions. This work was supported by the Estonian Research Council grants PRG803, TEM-TA23, RVTT3 and RVTT7, TARISTU24-TK10, TARISTU24-TK3, and the Center of Excellence program TK202 `Fundamental Universe'. This study was also financed, in part, by the São Paulo Research Foundation (FAPESP), Brasil, process numbers  2022/10785-4, and 2023/13275-0.  B. L. S\'anchez-Vega thanks the National Council for Scientific and Technological Development of Brazil, CNPq, for the financial support through grant n$^{\circ}$ 311699/2020-0.

\newpage
\bibliographystyle{apsrev4-1}
\bibliography{3-3-1}

\end{document}